\documentclass[american,aps,prb,reprint]{revtex4-1}
\usepackage[T1]{fontenc}
\usepackage[utf8]{inputenc}
\usepackage{color}
\usepackage{babel}
\usepackage{units}
\usepackage{textcomp}
\usepackage{amsbsy}
\usepackage{amstext}
\usepackage{amssymb}
\usepackage{graphicx}
\usepackage[unicode=true,pdfusetitle,
 bookmarks=true,bookmarksnumbered=false,bookmarksopen=false,
 breaklinks=true,pdfborder={0 0 0},pdfborderstyle={},backref=false,colorlinks=true]
 {hyperref}
\hypersetup{
 allcolors=blue}

\makeatletter

\providecommand{\tabularnewline}{\\}

\makeatother

\begin{document}

\title{Theory of the carbon vacancy in $\boldsymbol{4H}$-SiC: crystal field
and pseudo Jahn-Teller effects}

\author{José Coutinho}
\email{jose.coutinho@ua.pt}

\selectlanguage{american}%

\address{Department of Physics and I3N, University of Aveiro, Campus Santiago,
3810-193 Aveiro, Portugal}

\author{Vitor J. B. Torres}

\address{Department of Physics and I3N, University of Aveiro, Campus Santiago,
3810-193 Aveiro, Portugal}

\author{Kamel Demmouche}

\address{Institut des Sciences, Centre Universitaire -Belhadj Bouchaib- Ain
Temouchent, Route de Sidi Bel Abbes, B.P. 284, 46000 Ain Temouchent,
Algeria}

\author{Sven Öberg}

\address{Department of Engineering Sciences and Mathematics, Luleå University
of Technology, SE-97187 Luleå, Sweden}
\begin{abstract}
The carbon vacancy in $4H$-SiC is a powerful minority carrier recombination
center in as-grown material and a major cause of degradation of SiC-based
devices. Despite the extensiveness and maturity of the literature
regarding the characterization and modeling of the defect, many fundamental
questions persist. Among them we have the shaky connection of the
EPR data to the electrical measurements lacking sub-lattice site resolution,
the physical origin of the pseudo-Jahn-Teller effect, the reasoning
for the observed sub-lattice dependence of the paramagnetic states,
and the severe temperature-dependence of some hyperfine signals which
cannot be accounted for by a thermally-activated dynamic averaging
between equivalent Jahn-Teller distorted structures. In this work
we address these problems by means of semi-local and hybrid density
functional calculations. We start by inventorying a total of four
different vacancy structures from the analysis of relative energies.
Diamagnetic states have well defined low-energy structures, whereas
paramagnetic states display metastability. The reasoning for the rich
structural variety is traced back to the filling of electronic states
which are shaped by a crystal-field-dependent (and therefore site-dependent)
pseudo-Jahn-Teller effect. From calculated minimum energy paths for
defect rotation and transformation mechanisms, combined with the calculated
formation energies and electrical levels, we arrived at a configuration-coordinate
diagram of the defect. The diagram provides us with a detailed first-principles
picture of the defect when subject to thermal excitations. The calculated
acceptor and donor transitions agree well with the binding energies
of electrons emitted from the Z$_{1/2}$ and EH$_{6/7}$ traps, respectively.
From the comparison of calculated and measured $U$-values, and correlating
the site-dependent formation energies with the relative intensity
of the DLTS peaks in as-grown material, we assign Z$_{1}$ (EH$_{6}$)
and Z$_{2}$ (EH$_{7}$) signals to acceptor (donor) transitions of
carbon vacancies located on the $h$ and $k$ sub-lattice sites, respectively
\end{abstract}

\maketitle

\section{Introduction}

A wide and indirect band-gap, high chemical and thermal stability,
as well as radiation and electrical hardness, are among the merits
that make silicon carbide (SiC) an outstanding material for high-voltage
and high-power electronics.\cite{park1998,kimoto2014} Due to its
superior properties, the $4H$-SiC polytype has been the material
of choice of the industry. However, the presence of carbon-related
point defects in SiC, particularly carbon vacancies (V$_{\mathrm{C}}$),
is a major cause for minority carrier recombination in n-type material
and device failures like low field effect mobility.\cite{iwamoto2015,nipoti2015}
These problems have been connected to a set of V$_{\mathrm{C}}$-related
deep traps measured by deep-level transient spectroscopy (DLTS) and
labelled Z$_{1/2}$ and EH$_{6/7}$.\cite{kimoto1995,hemmingsson1997,hiyoshi2009}

The Z$_{1/2}$ has been ascribed to the superposition of Z$_{1}$
and Z$_{2}$ signals, each of which arising from a $\mathrm{V_{C}(=\!/0)}$
two-electron emission cascade at distinct sub-lattice sites of the
$4H$ polytype.\cite{hemmingsson1998,hemmingsson1999,son2012} Defects
behaving like that are said to possess a negative-$U$ as they show
an \emph{inverted order} of energy levels.\cite{watkins1984} This
is possible thanks to a strong atomic relaxation somewhere along the
emission sequence. For the case of Z$_{1/2}$ this translates into
the appearance of a $(=\!/0)$ occupancy level at about $E_{\mathrm{c}}-0.6$~eV,
implying that the formation of negatively charged vacancies ($\mathrm{V}{}_{\mathrm{C}}^{-}$)
is always energetically unfavorable against the formation of any mix
of neutral ($\mathrm{V}{}_{\mathrm{C}}^{0}$) and double negative
($\mathrm{V}{}_{\mathrm{C}}^{=}$) defects, no matter the position
of the Fermi energy. The appearance of $\mathrm{V}{}_{\mathrm{C}}^{-}$
is most likely when the Fermi level lies at the $(=\!/0)$ transition
energy, where its formation energy, $E_{\mathrm{f}}$, is lowest with
respect to other charge states. Depending on the temperature and the
energy difference $2E_{\mathrm{f}}(\mathrm{V_{C}^{-}})-E_{\mathrm{f}}(\mathrm{V_{C}^{=}}+\mathrm{V_{C}^{0}})$,
some $\mathrm{V}{}_{\mathrm{C}}^{-}$ states can still be populated.
Alternatively, $\mathrm{V}{}_{\mathrm{C}}^{-}$ can be formed from
other charge states after capture/emission of carriers upon optical
excitation. The actual $\mathrm{Z_{1}(=/-)}$ and $\mathrm{Z_{1}(-/0)}$
levels were respectively measured at 0.67~eV and $\sim0.52$~eV
below $E_{\mathrm{c}}$, whereas $\mathrm{Z_{2}(=/-)}$ and $\mathrm{Z_{2}(-/0)}$
were found at about $E_{\mathrm{c}}-0.71$~eV and $\sim E_{\mathrm{c}}-0.45$~eV,
respectively.\cite{hemmingsson1998,hemmingsson1999,son2012} Also
noteworthy is the fact that in $6H$-SiC, a pair of electron traps
located at $\sim E_{\mathrm{c}}-0.4$~eV and labelled E$_{1}$/E$_{2}$
from DLTS measurements, were attributed to acceptor transitions from
equivalent defects at different sub-lattice sites.\cite{dmowski1990,aboelfotoh1999}
More recently, high-resolution Laplace-DLTS was able to further resolve
E$_{1}$/E$_{2}$ into three components, and based on their similarity
with Z$_{1/2}$ (including their capture cross section and negative-$U$
ordering of levels), they were assigned to the carbon vacancy located
on all three available sites ($h$, $k_{1}$ and $k_{2}$) of the
$6H$ polytype.\cite{koizumi2013}

The EH$_{6/7}$ DLTS band has been a subject of discussion and surrounded
by some controversy. It usually shows up with a magnitude lower than
Z$_{1/2}$,\cite{ayedh2014} and it is made of two nearly overlapping
peaks, apparently with varying amplitude ratio (between 1:3 and 1:5)
depending on sample conditions.\cite{danno2006,booker2016} These
facts led to suggestions that EH$_{6/7}$ should not have the same
origin of Z$_{1/2}$, but rather be connected to a complex involving
V$_{\mathrm{C}}$.\cite{storasta2004,reshanov2007,zippelius2012,alfieri2013}
Recently, Booker and co-workers\cite{booker2016} analyzed the EH$_{6/7}$
capacitance transients, and based on a three-charge state model they
concluded that like Z$_{1/2}$, the EH$_{6/7}$ band results from
two correlated, two-electron emission processes from two defects.
Most importantly, they found that the concentration ratio of EH$_{6}$:EH$_{7}$
is 1:1 if we consider that the stronger peak actually combines EH$_{7}(0/+)$,
EH$_{7}(+/\!+\!+)$ and EH$_{6}(+/\!+\!+)$ transitions, while the
smaller component of the band comes from EH$_{6}(0/+)$ alone. The
issue of the inconsistent magnitude ratio between EH$_{6/7}$ and
Z$_{1/2}$ was poorly addressed. For all these transitions, carrier
binding energies were measured at $E_{\mathrm{c}}-1.50$~eV, $E_{\mathrm{c}}-1.46$~eV,
$E_{\mathrm{c}}-1.48$~eV and about $E_{\mathrm{c}}-1.42\textrm{-}1.49$~eV,
respectively. This suggests that EH$_{7}$ is a negative-$U$ defect,
while that cannot be said for EH$_{6}$ due to uncertainty in the
measurements.

\begin{figure}
\includegraphics[width=8.5cm]{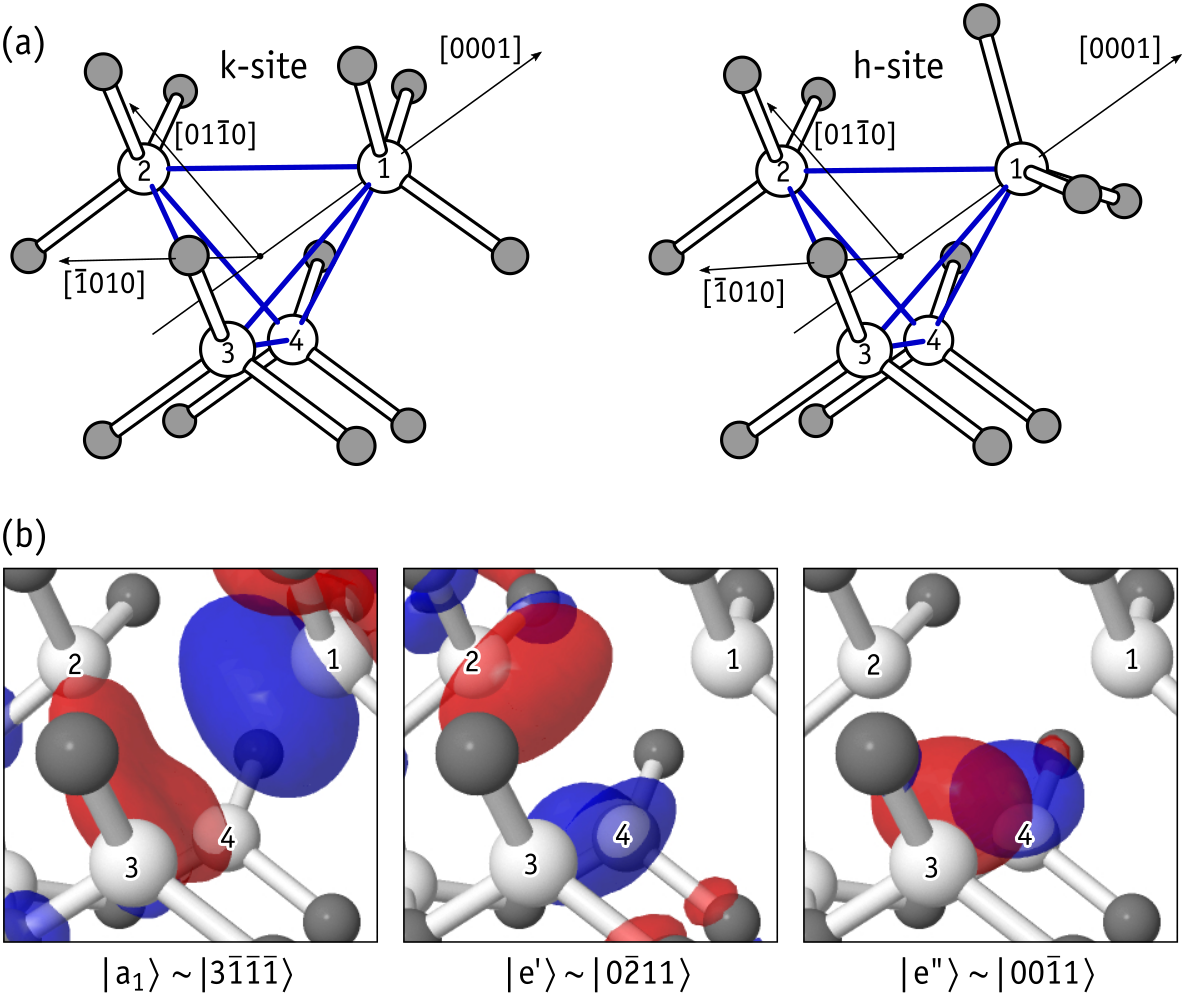}

\caption{\label{fig1}(a) Atomic structure of perfect carbon vacancies at $k$-
and $h$-sites of a $4H$-SiC crystal. Si and C atoms are white and
gray, respectively. Si$_{1}$ is axial (located on the {[}0001{]}
crystallographic axis) whereas Si$_{2\textrm{-}4}$ atoms lie on the
basal plane. (b) Representation of the one-electron states in the
gap ($a_{1}+e$) arising from an undistorted V$_{\mathrm{C}}$ defect
at the $k$-site. Red and blue isosurfaces denote negative and positive
phases, respectively.}
\end{figure}

Before continuing, let us introduce some notation with the help of
Figure~\ref{fig1}(a). Here we depict the atomic structure of perfect
V$_{\mathrm{C}}$ defects at $k$- and $h$-sites (with $k$ and $h$
labels referring to quasi-cubic and quasi-hexagonal sub-lattice sites
of the $4H$-SiC crystal). For the sake of convenience, the atom numbering
scheme was chosen in line with previous works in the literature.\cite{bockstedte2003,umeda2004b}
Hence, for a trigonal structure we have Si$_{1}$ (axial) and Si$_{2\textrm{-}4}$
(basal) shells of Si atoms. For monoclinic structures we assume Si$_{1}$
and Si$_{2}$ to lie on the $(\bar{1}010)$ mirror plane and the Si$_{3,4}$
pair to be mirror-symmetric. Hereafter, V$_{\mathrm{C}}^{q}(s)$ refers
to the carbon vacancy at the sub-lattice site $s\in\{k,\,h\}$ and
charge state $q\in\text{\{}=,\,-,\,0,\,+,\,+\!+\}$ (from double minus
to double plus). Occasionally we may also distinguish a vacancy with
a specific atomic geometry $R$ as V$_{\mathrm{C}}^{q}(s,R)$. We
also introduce at this point a way to represent the vacancy electronic
states using simple linear combination of atomic orbitals (LCAO).
Accordingly, a state $|\alpha_{1}\alpha_{2}\alpha_{3}\alpha_{4}\rangle$
stands for $A\sum_{i}\alpha_{i}\phi_{i}$, where $A$ is a normalization
constant, $\alpha_{i}$ are hybridization coefficients, and the summation
runs over all four Si$_{i}$ radical states $\phi_{i}$ (with $i=1,\ldots,4$).

Many details about the electronic and atomistic structure of V$_{\mathrm{C}}$
in $4H$-SiC, particularly in their paramagnetic $\mathrm{V_{C}^{+}}$
and $\mathrm{V_{C}^{-}}$ states, could be unraveled by electron paramagnetic
resonance (EPR) measurements.\cite{son2001,zvanut2002,konovalov2003,umeda2004a,bratus2005,umeda2005,son2012,trinh2013}
Among these reports, those combining experiments with first-principles
calculations\cite{umeda2004b,bratus2005,umeda2005,trinh2013} turn
out to be particularly elucidating. Below we provide a brief summary
of those results, with a special focus on the relevant issues for
the purpose of this work.

In p-type material irradiated with MeV electrons at high temperatures
(850$^{\circ}$C), the EPR spectrum revealed two signals, labelled
as EI5 (also referred to as Ky1/Ky2/ID1) and EI6 (also Ky3/ID2), which
were assigned to $V_{\mathrm{C}}^{+}(k)$ and $V_{\mathrm{C}}^{+}(h)$,
respectively.\cite{zvanut2002,konovalov2003,umeda2004a,umeda2004b,bratus2005}
Below $T\approx50$~K the main line of $V_{\mathrm{C}}^{+}(k)$ exhibited
$C_{1h}$ symmetry, and was accompanied by three distinct hyper-fine
(HF) signals due to interactions between the electron spin and $^{29}$Si
nuclei in shells with 1, 1 and 2 atoms. Above 50~K the spectrum was
converted to a trigonal ($C_{3v}$) pattern with two HFs representative
of 1 axial Si atom and 3 equivalent Si atoms on the basal plane. From
the temperature dependence of the HF life-times, the conversion from
monoclinic to trigonal symmetry was estimated to be limited by a barrier
as low as 0.014~eV.\cite{umeda2004b} $V_{\mathrm{C}}^{+}(h)$ on
the other hand, always showed trigonal symmetry irrespectively of
the temperature of the measurement (down to $T=4$~K). The HF structure
consisted of two line pairs with about 1:3 intensity ratio when the
magnetic field was aligned along {[}0001{]}. However, unlike for V$_{\mathrm{C}}^{+}$
at the cubic site, the HF principal direction of the basal radicals
of V$_{\mathrm{C}}^{+}(h)$ strongly deviated from the perfect tetrahedral
angle, and shifted from 103 down to 98\textdegree{} as the temperature
was lowered from 150~K to 10~K. This behavior was interpreted as
an increase of the anti-bonding character between the axial and basal
radicals when the temperature was lowered.\cite{umeda2004a}

Most observations described above were accounted for by density-functional
calculations. They arrived at ground state structures and HF tensors
compatible with the low-temperature EPR data.\cite{bockstedte2003,umeda2004b,bratus2005}
According to the calculations, V$_{\mathrm{C}}^{+}(k)$ and V$_{\mathrm{C}}^{+}(h)$
defects adopt $C_{1h}$ and $C_{3v}$ geometries in the ground state,
with their highest (semi-)occupied Kohn-Sham states (HOKS) possessing
$a'$ and $a_{1}$ symmetry, respectively. Within the above LCAO picture
they can be approximately described as $|a'\rangle\sim|11\bar{1}\bar{1}\rangle$
and $|a_{1}\rangle\sim|3\bar{1}\bar{1}\bar{1}\rangle$, respectively,
explaining the HF structure observed for V$_{\mathrm{C}}^{+}(k)$
and V$_{\mathrm{C}}^{+}(h)$ at low-temperatures ($\sim5$~K). They
are also consistent with the measurements of V$_{\mathrm{C}}^{+}(k)$
at $T>50$~K if we assume that above this temperature the defect
assumes a motional-averaged trigonal state due to fast hopping between
all three $|11\bar{1}\bar{1}\rangle$, $|1\bar{1}1\bar{1}\rangle$
and $|1\bar{1}\bar{1}1\rangle$ degenerate structures neighboring
the undistorted ($C_{3v}$) configuration. Note that in line with
the observations, all Si radicals contribute to $|a'\rangle$ in V$_{\mathrm{C}}^{+}(k)$
(under static and dynamic conditions), and the amplitude of the axial
radical $\phi_{1}$ in the $|a_{1}\rangle$ state accounts for about
50\% of the total LCAO localization in V$_{\mathrm{C}}^{+}(h)$. The
$|a_{1}\rangle$ paramagnetic state of V$_{\mathrm{C}}^{+}(h)$ is
also consistent with the observed anti-bonding character between Si$_{1}$
and the basal Si$_{2\textrm{-}4}$ radicals. However, the model is
still unable to account for the magnitude of the high-temperature
($T>30$~K) HF signals. Another puzzle, which was noted by Bockstedte
and co-workers,\cite{bockstedte2003} is that despite being a singlet
state, the trigonal V$_{\mathrm{C}}^{+}(k)$ configuration is unstable
against monoclinic distortion, implying the influence of a pseudo-Jahn-Teller
(pJT) effect. However, neither was a justification provided for its
manifestation, nor was it found why a similar effect is apparently
missing in V$_{\mathrm{C}}^{+}(h)$.

Negatively charged carbon vacancies were observed by EPR in n-type
$4H$-SiC irradiated either with MeV electrons at 850$^{\circ}$C
or with 250~keV electrons at room temperature.\cite{umeda2005,son2012}
Although some traces of V$_{\mathrm{C}}^{-}(k)$ and V$_{\mathrm{C}}^{-}(h)$
signals could be detected above $T\approx100$~K in darkness (in
heavily doped material),\cite{son2012} most experiments were performed
on illuminated samples, which gave rise to much stronger signals.\cite{umeda2005,son2012,trinh2013}
At the cubic site and below $T\approx40$~K, the V$_{\mathrm{C}}^{-}(k)$
main line showed a monoclinic pattern and a single HF pair related
to two symmetry-equivalent Si nuclei (Si$_{3,4}$).\cite{trinh2013}
Additional and weaker HF signals were related to more distant shells.
Above $T\approx40$~K the Si$_{3,4}$ HF signal disappeared from
the spectrum and the main line acquired a trigonal pattern, accompanied
by the appearance of a new axial HF pair (due to interaction between
a magnetic Si$_{1}$ nucleus and the electron spin). As the temperature
further increased, the magnitude of the Si$_{1}$ HF splitting increased
and at $\sim$80-90~K a weak and broad HF pair accounting for three
equivalent (Si$_{2\textrm{-}4}$) nuclei appeared in the spectrum
as well.\cite{trinh2013} Regarding V$_{\mathrm{C}}^{-}(h)$, the
main signal is monoclinic at $T=60$~K and below. At these temperatures
two HF signal pairs related to two inequivalent Si nuclei (Si$_{1}$
and Si$_{2}$) with $C_{1h}$ site symmetry were detected. Raising
the temperature above $T\sim70$~K led to the broadening and disappearance
of the Si$_{2}$ HF, while the main-line and Si$_{1}$ HF components
merged into single trigonal peaks. The activation barrier for the
monoclinic-trigonal conversion was estimated as 0.02~eV.\cite{umeda2005}
Within the temperature range of 70-120~K only the Si$_{1}$ HF was
detected, but when $T>120$~K a trigonal HF signal representative
of three equivalent Si nuclei (Si$_{2\textrm{-}4}$) was also observed.

Again, first-principles modeling played a key role in grasping several
of the above features.\cite{zywietz1999,umeda2005,trinh2013} Recent
density functional calculations indicated that V$_{\mathrm{C}}^{-}(k)$
has a $|a''\rangle\sim|00\bar{1}1\rangle$ paramagnetic ground state,
whereas the symmetric $|a'\rangle\sim|1\bar{1}00\rangle$ state was
metastable by only 0.03~eV.\cite{trinh2013} The calculated HF tensors
for Si$_{3,4}$ radicals accounted very well for the low-temperature
($T=30$~K) experimental data. The quenching of the Si$_{3,4}$ HF
signal above 40~K and the observation of trigonal hyperfine structures
at higher temperatures was suggested to result from the partial population
of both $|a''\rangle$ and $|a'\rangle$ states. Accordingly, under
these conditions they would quickly hop between three equivalent Jahn-Teller
(JT) distorted alignments. While this picture aims at accounting for
the observed non-zero amplitude of the wave-function on all four radicals
above 90~K, it cannot be correct. Any sequential transformation between
$|a'\rangle$ and $|a''\rangle$ states involves an intermittent quenching
of the spin-density on the basal nuclei. Further, the model could
not explain why there is a $\sim40$~K gap between the quenching
of the Si$_{3,4}$ HF signal (at 40~K) and the appearance of the
Si$_{2\textrm{-}4}$ HF signal (at 80~K). Also puzzling and unexplored
was the fact that the V$_{\mathrm{C}}^{-}(k)$ ground state was found
to be nodal ($a''$), which in principle has higher kinetic energy
than the metastable state ($a'$). Finally, the symmetry lowering
of V$_{\mathrm{C}}^{-}(k)$ cannot simply be explained by the JT effect.
In the perfect vacancy ($C_{3v}$ symmetry), the four Si$_{1\textrm{-}4}$
radicals hybridize into a fully occupied valence state $|a_{1}\rangle\sim|1111\rangle$,
and three gap states $|a_{1}\rangle\sim|3\bar{1}\bar{1}\bar{1}\rangle$,
$|e'\rangle\sim|0\bar{2}11\rangle$ and $|e''\rangle\sim|00\bar{1}1\rangle$
to be populated with three electrons. We calculated these states for
an undistorted (trigonal) vacancy at the $k$-site using the same
method of Ref.~\onlinecite{trinh2013} and they are depicted in Figure~\ref{fig1}(b).
The latter two are higher in energy and represent orthogonal components
of a doublet which is split from $|a_{1}\rangle$ due to the internal
crystal field. For the case of V$_{\mathrm{C}}^{-}(k)$ the doublet
becomes partially populated (with a single electron) and the JT effect
is expected to split $(e''+e')^{\uparrow}$ (within $C_{3v}$) into
$(a''^{\uparrow}+a')$ (within $C_{1h}$), where the upward arrow
stands for the paramagnetic electron. Now, while the first-principles
results from Ref.~\onlinecite{trinh2013} indicate that the metastable
$|a'\rangle$ state has amplitude on Si$_{1}$, it is clear from Figure~1(b)
that a JT-split component $|e'\rangle\sim|0\bar{2}11\rangle$ cannot
account for this feature.

Turning now to V$_{\mathrm{C}}^{-}(h)$, the calculations arrived
at a $C_{1h}$ ground state rather different than that found for the
cubic site, namely the unpaired electron was localized on the Si$_{1}$-Si$_{2}$
pair as $|a'\rangle\sim|1\bar{1}00\rangle$.\cite{zywietz1999,umeda2005}
The calculated HF tensor elements for both (inequivalent) Si$_{1}$
and Si$_{2}$ radicals agreed very well with the measurements below
$T=60$~K (both in magnitude and principal directions), providing
compelling evidence for the correctness of the model. The disappearance
of the Si$_{2}$ HF signal together with the conversion of the $C_{1h}$-symmetric
Si$_{1}$ HF into a trigonal signal at $T>70$~K was justified based
on a thermal activated hopping between $|1\bar{1}00\rangle$, $|10\bar{1}0\rangle$
and $|100\bar{1}\rangle$ equivalent states, which preserves a steady
wave function amplitude only on Si$_{1}$.\cite{umeda2005} Again,
the reasoning for a 70-120~K temperature window where only Si$_{1}$
HF was observed and above which another trigonal Si$_{2\textrm{-}4}$
HF was observed, was left unaddressed. Analogously to the metastable
structure in the cubic site, the electronic structure of V$_{\mathrm{C}}^{-}(h)$
in the ground state involves a non-vanishing spin-density on Si$_{1}$.
Hence, unlike suggested in Ref.~\onlinecite{umeda2005}, the model
cannot be explained by the JT effect, simply because none of the $e$-components
in Fig.~\ref{fig1}(b) shows non-zero amplitude on Si$_{1}$. Finally,
V$_{\mathrm{C}}^{-}(k)$ and V$_{\mathrm{C}}^{-}(h)$ show monoclinic
ground-states with opposite symmetry with respect to the mirror plane.
Although the calculations were successful in accounting for the observed
site-dependent ordering of electronic states,\cite{zywietz1999,umeda2005}
again the reasonings behind this effect were left unaddressed.

The connection of $\mathrm{V_{C}}$ (via EPR) with the Z$_{1/2}$
and EH$_{6/7}$ traps (via DLTS) was suggested based on the correlation
between the position of the DLTS levels and the photo-EPR excitation
thresholds for $\mathrm{V_{C}^{=}}\rightarrow\mathrm{V_{C}^{-}}+e^{-}$
and $\mathrm{V_{C}^{0}}\rightarrow\mathrm{V_{C}^{+}}+e^{-}$, respectively
(where $e^{-}$ represent a free electron at the conduction band bottom).\cite{son2012}
More recently, Kawahara \emph{et~al.}\cite{kawahara2013,kawahara2014}
investigated samples irradiated by low-energy (250~keV) electrons,
which could displace C atoms only. In those works they reported a
good correlation between the area density of EPR active $\mathrm{V_{C}^{-}}$
and the fraction of carriers trapped by the dominant Z$_{1/2}$ on
samples irradiated with different electron fluences.

It seems clear that Z$_{1/2}$ is a negative-$U$ center. This is
consistent with the need of optical excitation in order to observe
negatively charged vacancies by EPR. However, that is not the case
for the defect responsible for EH$_{6/7}$. In recent state-of-the-art
electrical level calculations using many-body perturbation\cite{bockstedte2010}
and hybrid density functional\cite{hornos2011} methods, the donor
levels were predicted to be separated by a small positive or essentially
zero $U$-value ($U\approx0.0\textrm{-}0.2$~eV). While this agrees
with the low-temperature EPR measurements in darkness, it is also
in apparent conflict with the negative-$U$ ordering reported for
EH$_{7}$ and tentatively proposed for EH$_{6}$ from DLTS.\cite{booker2016}
As a word of caution, we note that when periodic charge corrections
were neglected, the calculations clearly indicated $U<0$ for both
acceptors and donors.\cite{zywietz1999,bockstedte2010,trinh2013}

It is clear that despite many advances, there are several fundamental
puzzles to be solved. This paper aims at addressing those issues,
as well as others that will become evident further ahead. In this
section we wanted to introduce the reader to the main properties of
the carbon vacancy in $4H$-SiC, how the EPR data has been related
to the prominent Z$_{1/2}$ and EH$_{6/7}$ electron traps, and the
importance of theory/computational modeling in providing models and
checking their quality. We will now proceed with a description of
the theoretical methods followed by the main results. These include
the reproduction of structures and electronic levels previously reported,
as well as new results like a physical description ofthe observed
pseudo-Jahn-Teller distortions, the crystal-field impact on the distinct
electronic structure of cubic and hexagonal vacancies, and the atomistic
mechanisms behind the $T$-dependent dynamic effects observed by EPR.
Before the conclusions, we also include a section where we discuss
the above issues.

\section{Theory}

The calculations were carried out using the VASP package,\cite{kresse1993,kresse1994,kresse1996a,kresse1996b}
employing the projector-augmented wave (PAW) method to avoid explicit
treatment of core electrons.\cite{blochl1994} A plane wave basis
set with kinetic energy up to 400~eV was used to describe the electronic
Kohn-Sham states. The many-body electronic potential was evaluated
using the hybrid density functional of Heyd-Scuseria-Ernzerhof (HSE06),\cite{heyd2003,krukau2006}
which mixes semi-local and exact exchange interactions at short ranges,
treating the long-range interactions within the simpler generalized
gradient approximation as proposed by Perdew, Burke and Ernzerhof
(PBE).\cite{perdew1996} When compared to plain PBE calculations,
HSE06 has the main advantage of predicting a Kohn-Sham (indirect)
band gap 3.17~eV wide for $4H$-SiC, which should be compared to
the experimental value of 3.27~eV.\cite{grivickas2007} To a large
extent, this approach mitigates the well known underestimated gap
\emph{syndrome} affecting PBE-level calculations, which show a 2.19~eV
band gap width. Although most results reported below were obtained
using the HSE06 method, PBE-level results are also included and in
that case they are explicitly identified.

We used 576-atom hexagonal supercells, obtained by replication of
$6\!\times\!6\!\times\!2$ unit cells, from where a carbon atom was
removed to produce a V$_{\mathrm{C}}$ defect. The equilibrium (calculated)
lattice parameters of $4H$-SiC were $a=3.071$~Å and $c=10.05$2~Å.
These are close to the experimental values of $a=3.081$~Å and $c=10.085$~Å
extrapolated to $T=0$~K.\cite{li1986} All defect structures were
optimized within PBE-level, using a conjugate-gradient method until
the forces acting on the atoms were lower than 10~meV/Å. After this
step, we took the relaxed structure, and self-consistent energies,
electron and spin densities were finally obtained within HSE06. Electronic
relaxations were computed with a numerical accuracy of 1~$\mu\mathrm{eV}$,
and the band structures were solved at $\mathbf{k}=(0\,0\,\nicefrac{1}{2})$
in reciprocal lattice units. This is conventionally referred to as
the $A$-point in the hexagonal Brillouin zone (BZ). We found this
particular $\mathbf{k}$-point to provide the best compromise between
sampling accuracy and computational performance. It is representative
of the $\mathbf{k}$-point set $(0\,0\,\pm\!\nicefrac{1}{2})$ in
non-relativistic calculations, it led to energy differences with an
error bar of about 5~meV (when compared to results obtained using
$2\!\times\!2\!\times\!2$ sampled BZ), and most importantly, it did
not cause so strong hybridization between defect levels lying high
in the gap and the conduction band states as in the $\Gamma$-sampled
PBE calculations of Ref.~\onlinecite{trinh2013}.

The above two-step recipe to obtain hybrid density-functional energies
using structures that were previously relaxed within PBE (hereafter
referred to as \emph{pseudo-relaxed energies}), casts doubts regarding
its accuracy when compared to \emph{fully-relaxed} HSE06-energies
obtained by minimizing HSE06-forces. To clarify this issue, we compared
energies and forces of pseudo- and fully-relaxed V$_{\mathrm{C}}^{+\!+}(k)$
and V$_{\mathrm{C}}^{=}(k)$ states. These two charge states have
rather different structures (to be discussed below), and while V$_{\mathrm{C}}^{+\!+}(k)$
does not have electrons occupying gap levels, V$_{\mathrm{C}}^{=}(k)$
has two fully occupied gap states, one of them being close to the
conduction band edge. These tests were carried out using 256-atom
supercells ($4\!\times\!4\!\times\!2$ unit cells) with a $\Gamma$-centered
grid of $2^{3}$ $\mathbf{k}$-points for BZ sampling. Accordingly,
we obtained fully-relaxed HSE06-energies 13~meV and 11~meV below
the energy of pseudo-relaxed V$_{\mathrm{C}}^{+\!+}(k)$ and V$_{\mathrm{C}}^{=}(k)$
calculations, respectively. Despite these small relaxation energies,
the average HSE06-force acting on Si$_{1\textrm{-}4}$-atoms on PBE-relaxed
structures were 0.28~eV/Å and 0.10~eV/Å for for V$_{\mathrm{C}}^{+\!+}(k)$
and V$_{\mathrm{C}}^{=}(k)$, respectively, and therefore cannot be
neglected. However, the energy difference $E(q=-2)-E(q=+2)$ was 39.384~eV
and 39.386~eV for pseudo-relaxed and fully-relaxed calculations,
respectively, suggesting that the error of pseudo-relaxed energy differences
is of the order of a few meV. 

The energy of a charged defect, when calculated using periodic boundary
conditions, is actually the energy of a supercell contaminated by
artificial Coulomb interactions across an array of charged defects
embedded on a compensating background charge.\cite{makov1995} These
interactions are long-ranged and difficult to remove. Several \emph{post-processing}
recipes have been proposed to mitigate this problem (see for example
Ref.~\onlinecite{komsa2012} and references therein). Here we use
the method by Freysoldt, Neugebauer and Van de Walle,\cite{freysoldt2009}
recently generalized for anisotropic materials.\cite{kumagai2014}
Accordingly, the total energy of a defect in an \emph{infinite} crystal
is approximately $E_{\mathrm{def}}(q)\approx\tilde{E}_{\mathrm{def}}(q)+E_{\mathrm{corr}}(q)$,
where $q$ is a localized net charge on the defect, $\tilde{E}_{\mathrm{def}}$
is the total energy of the periodic problem and $E_{\mathrm{corr}}$
the charge correction,

\begin{equation}
E_{\mathrm{corr}}(q)=E_{\mathrm{PC}}(q)-q\Delta\bar{\phi}_{\mathrm{PC,ind}}(q),\label{eq:e-corr}
\end{equation}
where $E_{\mathrm{PC}}(q)$ is a point charge correction, which for
isotropic materials reduces to the Madelung energy $E_{\mathrm{PC,iso}}(q)=\alpha_{\mathrm{M}}q^{2}/2\epsilon L$
and depends on the ratio between the Madelung constant $\alpha_{\mathrm{M}}$
and a characteristic length $L$ (usually a lattice constant), the
net charge and the dielectric constant $\epsilon$. Further details
about the explicit calculation of $E_{\mathrm{PC}}(q)$ for anisotropic
materials (like $4H$-SiC) can be found in Ref.~\onlinecite{kumagai2014}.

\begin{equation}
\Delta\bar{\phi}_{\mathrm{PC,ind}}(q)=\bar{\phi}_{\mathrm{ind}}(q)-\bar{\phi}_{\text{PC}}(q)
\end{equation}
is the offset between the defect induced average potential $\bar{\phi}_{\mathrm{ind}}(q)=\bar{\phi}_{\mathrm{def}}(q)-\bar{\phi}_{\mathrm{bulk}}$
and that produced by a point-charge, $\bar{\phi}_{\mathrm{PC}}(q)$.\cite{kumagai2014}
The space-averaged potentials $\bar{\phi}_{\mathrm{def}}$ and $\bar{\phi}_{\mathrm{bulk}}$
are obtained from first-principles from the Hartree (electrostatic)
potential considering defective and pristine (bulk) supercells. The
averaging is done at remote locations from the defect, more precisely
at all atomic sites outside the largest sphere inscribed by the Wigner-Seitz
supercell (see Figure~2(a) of Ref.~\onlinecite{kumagai2014}). For
the 576-atom supercell employed in this work, that meant a 15.1~Å
radius sphere leaving a total of 382 outer atomic sites to be sampled.

Since defects distort and polarize the surrounding lattice, besides
the electronic (ion-clamped) component, $\boldsymbol{\epsilon}_{\infty}$,
the dielectric tensor employed in the calculation of $E_{\mathrm{corr}}$
has to account for the ionic contribution as well, $\boldsymbol{\epsilon}=\boldsymbol{\epsilon}_{\infty}+\boldsymbol{\epsilon}_{\mathrm{ion}}$.\cite{komsa2012}
We calculated $\boldsymbol{\epsilon}_{\infty}$ using density-functional
perturbation theory with local field effects within PBE.\cite{baroni1986}
The ionic part was evaluated from the Born effective charges and eigen-frequencies
of the Hessian matrix.\cite{cockayne2000} Accordingly, we obtained
$\epsilon^{\parallel}=10.65$ and $\epsilon^{\perp}=9.88$ for the
dielectric constant parallel and perpendicular to the crystallographic
$c$-axis. These figures account well for the values $\epsilon^{\parallel}=10.03$
and $\epsilon^{\perp}=9.66$ obtained from refractive index measurements
and Raman scattering data.\cite{patrick1970}

\begin{figure}
\includegraphics[width=8.3cm]{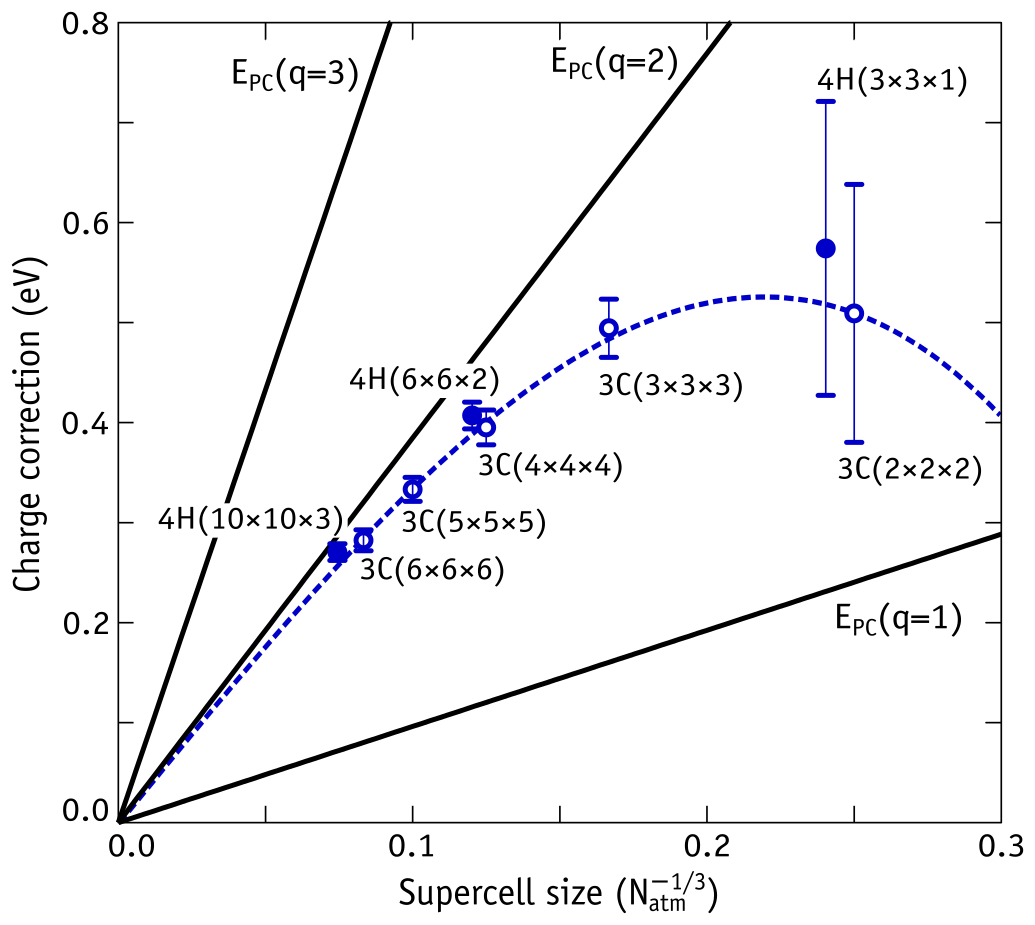}

\caption{\label{fig2}Variation of the charge correction, $E_{\mathrm{corr}}$,
obtained for a double positively charged carbon vacancy in $4H$-SiC
(closed circles) and $3C$-SiC (open circles) supercells sized by
the number of atoms, $N_{\mathrm{atm}}$. Error bars are standard
deviations obtained from the averaging of $\Delta\bar{\phi}_{\mathrm{PC,ind}}$
(see Eq.~\ref{eq:e-corr}). Integer triplets $(l\times m\times n)$
are scaling factors applied to the lattice vectors of each unit cell
to obtain the respective supercell. The dashed line represents a function
of the form $aN_{\mathrm{atm}}^{-1}+bN_{\mathrm{atm}}^{-1/3}$ fitted
to the data. Solid lines represent simple point charge corrections
for $q=1,\,2$ and 3.}
\end{figure}

Figure~\ref{fig2} depicts the values of $E_{\mathrm{corr}}$ obtained
for a $V_{\mathrm{C}}^{+\!+}$ defect in $4H$-SiC (with $C_{3v}$
symmetry) and $3C$-SiC (with $T_{d}$ symmetry) as a function of
the number of atoms in the supercell ($N_{\mathrm{atm}}\sim L^{3}$).
The calculations shown in this particular figure were carried out
within PBE-level. The results were essentially the same when using
the HSE06 functional. Integer triplets in the figure $(l\times m\times n)$
are scaling factors applied to the lattice vectors of the conventional
cell (8 atoms in both polytypes) to obtain the respective supercell.
For example, the largest hexagonal cell ($4H$-SiC) consisted of $(10\times10\times3)\times8=2400$~atoms,
whereas the smallest cubic cell ($3C$-SiC) had $(2\times2\times2)\times8=64$~atoms.
$E_{\mathrm{corr}}$ can be expanded in a power series of $L$, with
the first term $E_{\mathrm{PC}}\sim L^{-1}$ and the second term scaling
as $L^{-3}$.\cite{makov1995,castleton2006} The data were therefore
used to fit a function of the form $aN_{\mathrm{atm}}^{-1}+bN_{\mathrm{atm}}^{-1/3}$,
which is shown by the dashed line. The solid straight lines represent
the leading term, $E_{\mathrm{PC}}(q=1,2,3)$, as a function of $N_{\mathrm{atm}}$.
It is clear that the simple Madelung (point-like) correction overestimates
the spurious Coulomb energy. Also as expected, $E_{\mathrm{corr}}$
asymptotically converges to $E_{\mathrm{PC}}$ for supercells of infinite
size. The error of the Madelung correction is always above the statistical
error of $E_{\mathrm{corr}}$ obtained from averaging $\Delta\bar{\phi}_{\mathrm{PC,ind}}$
and shown as error bars. In the case of the $4H$-SiC $(6\times6\times2)$
supercells (used in this work to study the carbon vacancy), our best
estimate for the correction of $V_{\mathrm{C}}^{+\!+}$ is $E_{\mathrm{corr}}(q=2)=0.41\pm0.01$~eV,
whereas for a singly charged vacancy (not shown in the graph) we obtained
$E_{\mathrm{corr}}(q=1)=0.10\pm0.01$~eV.

For the calculation of formation energies we follow the usual procedure,
introduced by Qian, Martin and Chadi.\cite{qian1988} Here the formation
energy of a carbon vacancy is

\begin{equation}
E_{\mathrm{f}}(q,R,\mu_{\mathrm{C}},E_{\mathrm{F}})=E_{\mathrm{def}}(q,R)-E_{\mathrm{bulk}}+\mu_{\mathrm{C}}+q(E_{\mathrm{v}}+E_{\mathrm{F}}),\label{eq:formation}
\end{equation}
where $E_{\mathrm{def}}$ is the charge-corrected total energy of
the defective supercell as defined above. Besides depending on the
charge state $q$, $E_{\mathrm{def}}$ may refer to more than one
atomic structure $R$. $E_{\mathrm{bulk}}$ is the energy of a perfect
supercell, $\mu_{\mathrm{C}}$ is the carbon chemical potential (see
below), $E_{\mathrm{v}}$ is the valence band edge and $E_{\mathrm{F}}$
the Fermi energy which may vary within $E_{\mathrm{F}}=[0,E_{\mathrm{g}}]$.
The upper limit, $E_{\mathrm{g}}=I_{\mathrm{bulk}}-A_{\mathrm{bulk}}=3.42$~eV,
is the calculated forbidden gap width, here obtained within the delta
self-consistent ($\Delta$SCF) method,\cite{hedin1970} where $A_{\mathrm{bulk}}=E_{\mathrm{bulk}}(0)-E_{\mathrm{bulk}}(-1)=-11.03$~eV
and $I_{\mathrm{bulk}}=E_{\mathrm{bulk}}(+1)-E_{\mathrm{bulk}}(0)=-7.61$~eV
are ionization potentials of neutral and negatively charged supercells.
They are negative as their reference (zero-energy) is ill-defined
for a periodic calculation. According to this method $E_{\mathrm{v}}=-I_{\mathrm{bulk}}$,
allowing to consistently express the calculated transition levels
with respect to both $E_{\mathrm{c}}$ and $E_{\mathrm{v}}$ without
having to rely on the experimental band gap.

We note that using the $A$-point for BZ sampling, the $E_{\mathrm{g}}=3.42$~eV
obtained by the $\Delta$SCF method is 0.25~eV wider than the indirect
gap from the Kohn-Sham energies of the highest-occupied and lowest-unoccupied
states at $\mathbf{k}=\Gamma$ and $\mathbf{k}=M=(\nicefrac{1}{2}\,0\,0)$,
respectively. This compares with $E_{\mathrm{g}}=3.15$~eV and $E_{\mathrm{g}}=3.25$~eV
from analogous $\Delta$SCF calculations using $\Gamma$-centered
$1\times1\times1$ (simple $\Gamma$-point) and $2\times2\times2$
$\mathbf{k}$-point sampling grids. The $\Gamma$-sampled $E_{\mathrm{g}}$
energy coincides with the Kohn-Sham gap simply because of band-folding,
which for a $6\times6\times2$-supercell brings the $M$-point into
the origin of the BZ. These results indicate that $1\times1\times1$-sampled
energy differences (like $\Gamma$- and $A$-point calculations) may
suffer from insufficient sampling density. This effect is expected
to be more severe for energy differences involving the occupation
(or emptying) of highly dispersive states. The calculation of $I_{\mathrm{bulk}}-A_{\mathrm{bulk}}$
is perhaps an extreme case. It involves emptying the top-most valence
band and filling the bottom-most conduction band, both showing considerable
dispersion amplitudes. On the other hand, for sufficiently large supercells,
localized defect states show little dispersion and sampling errors
tend to cancel when considering energy differences. This is confirmed
by $\Gamma$-point, $A$-point and $2\times2\times2$-grid calculations
of $E_{\mathrm{def}}(q=-2)-E_{\mathrm{def}}(q=+2)$ for the vacancy
at the $k$-site, which gives an average value and maximum deviation
of $39.480\pm0.003$~eV.

In Eq.~\ref{eq:formation}, $\mu_{\mathrm{C}}$ represents the energy
per carbon atom in the SiC crystal, which is subject to

\begin{equation}
\mu_{\mathrm{C}}^{0}+\Delta H_{\mathrm{SiC}}^{\mathrm{f}}\leq\mu_{\mathrm{C}}\leq\mu_{\mathrm{C}}^{0},\label{eq:mu_bounds}
\end{equation}
where the upper and lower bounds represent C-rich and C-poor SiC crystals,
which are in equilibrium with standard carbon and silicon phases,
respectively. For crystals grown under stoichiometric conditions we
have $\mu_{\mathrm{C}}=\mu_{\mathrm{C}}^{0}+\Delta H_{\mathrm{SiC}}^{\mathrm{f}}/2$.
Here $\Delta H_{\mathrm{SiC}}^{\mathrm{f}}$ is the heat of formation
of SiC estimated as $\Delta H_{\mathrm{SiC}}^{\mathrm{f}}=\mu_{\mathrm{SiC}}^{0}-\mu_{\mathrm{C}}^{0}-\mu_{\mathrm{Si}}^{0}=-0.62$~eV,
with $\mu_{\mathrm{SiC}}^{0}$ being the energy per SiC formula unit
in a perfect crystal, while $\mu_{\mathrm{C}}^{0}$ and $\mu_{\mathrm{Si}}^{0}$
are $standard$ chemical potentials (energy per atom) of C and Si
in diamond and silicon crystals, respectively. The value calculated
for $\Delta H_{\mathrm{SiC}}^{\mathrm{f}}$ is close to $-0.72$~eV
as obtained from calorimetry measurements.\cite{greenberg1970}

An important use of Eq.~\ref{eq:formation} is in locating the value
of $E_{\mathrm{F}}$ for which two different charge states, say $q$
and $q+1$, have the same energy, and therefore the same probability
to occur. The $(q/q+1)$ transition level with respect to the valence
band top is found at $E_{\mathrm{F}}=E(q/q+1)-E_{\mathrm{v}}$ such
that $E_{\mathrm{f}}(q,R_{q},\mu_{\mathrm{C}},E_{\mathrm{F}})=E_{\mathrm{f}}(q+1,R_{q\!+\!1},\mu_{\mathrm{C}},E_{\mathrm{F}})$,

\begin{equation}
E(q/q+1)-E_{\mathrm{v}}=\left[E_{\mathrm{def}}(q,R_{q})-E_{\mathrm{def}}(q+1,R_{q\!+\!1})\right]-I_{\mathrm{bulk}},\label{eq:donor}
\end{equation}
where we distinguish eventual different structures $R_{q}$ and $R_{q\!+\!1}$
for charge states $q$ and $q+1$, respectively. It is also useful
to calculate transition levels with respect to the conduction band
minimum. For that we have,

\begin{equation}
E_{\mathrm{c}}-E(q/q+1)=A_{\mathrm{bulk}}-\left[E_{\mathrm{def}}(q,R_{q})-E_{\mathrm{def}}(q+1,R_{q\!+\!1})\right].\label{eq:acceptor}
\end{equation}

We also investigated the transformation of V$_{\mathrm{C}}$ defects
between different structures and also between different symmetry-equivalent
alignments. We assume the adiabatic approximation, and the potential
energy surface governing the atomic motion was calculated using the
climbing-image nudged elastic band (NEB) method.\cite{henkelman2000}
The NEB algorithm allows to find saddle points and minimum energy
paths separating known initial and final structures. The method optimizes
a number of intermediate structures along the reaction path while
maintaining equal \emph{spacing} between them. This is possible thanks
to the introduction of spring forces connecting neighboring structures
(the elastic band) and projecting out the component of the force due
to the potential perpendicular to the band. The NEB relaxations were
carried out within the PBE-level, used 7 intermediate structures,
and the forces acting on the atoms were also converged within $0.01$~eV/Å.
The initial, final and saddle-point structures ($R_{\mathrm{i}}$,
$R_{\mathrm{f}}$ and $R_{\mathrm{sp}}$, respectively) were used
to obtain their respective total energies ($E_{\mathrm{i}}$, $E_{\mathrm{f}}$
and $E_{\mathrm{sp}}$) using the HSE06 functional.

\section{Results}

\begin{figure}
\includegraphics[width=8.3cm]{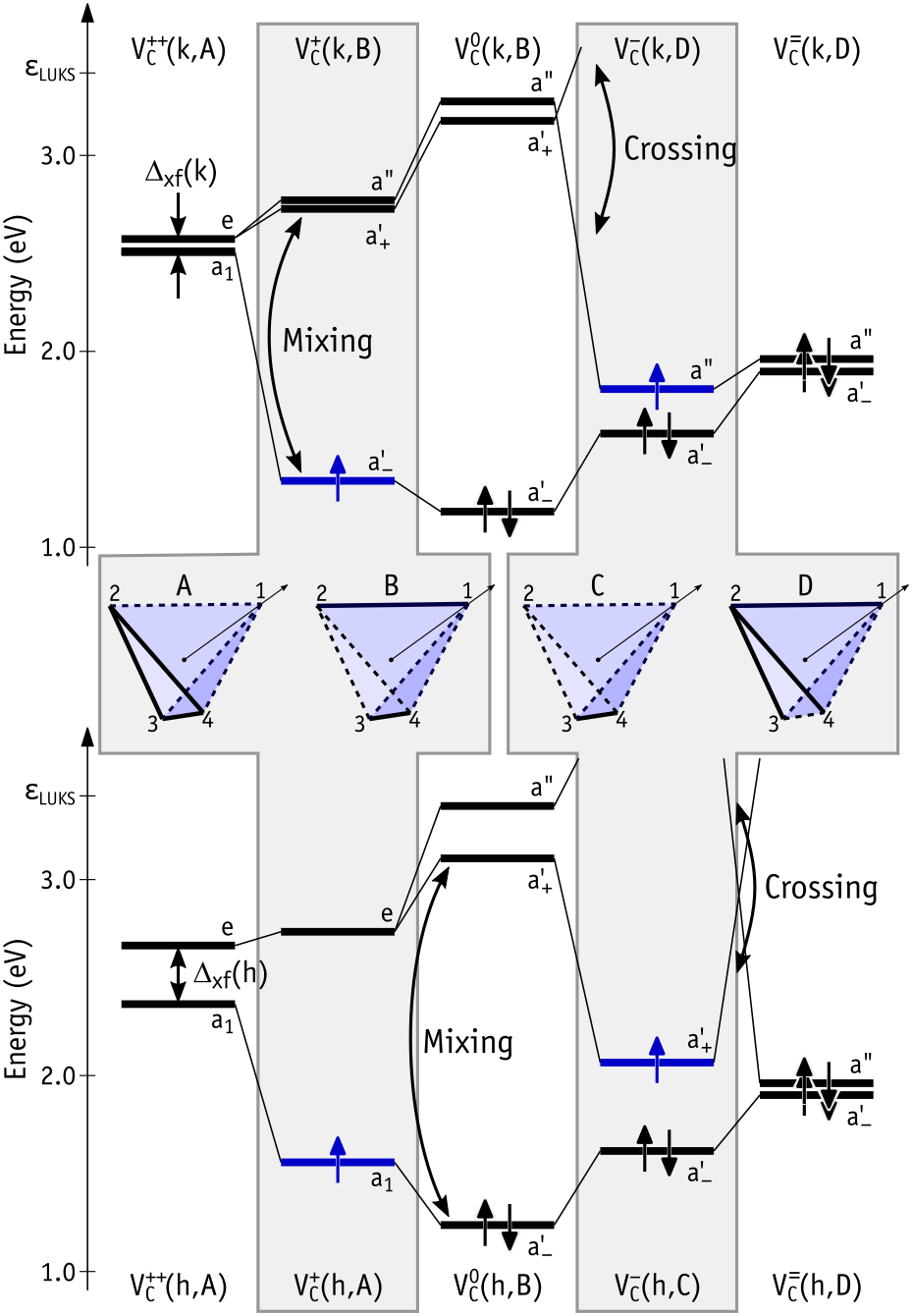}

\caption{\label{fig3}Kohn-Sham electronic states at $\mathbf{k}=(0\,0\,\nicefrac{1}{2})$
for the carbon vacancy at the cubic (upper half) and hexagonal (lower
half) sites. The zero of the energy scale is at the $\epsilon_{\mathrm{HOKS}}$
energy in bulk (using the HSE06 functional). Electrons are represented
by upward (spin-up) and downward (spin-down) arrows. Each state is
accompanied by symmetry labels (see text for details). The central
region displays the structures found for positively charged (A and
B) and negatively charged (C and D) defects. Contracted/elongated
Si-Si distances are represented as solid/dashed lines, respectively.}
\end{figure}

\subsection{Ground-state results for the carbon vacancy}

We start by reporting on the structural properties of the defect on
different charge states. The V$_{\mathrm{C}}$ defect was always found
to have the lowest energy in low-spin states. We identified four different
atomistic configurations, which we label with the letters A (with
$C_{3v}$ symmetry), and B, C and D (with $C_{1h}$ symmetry). They
are distinguished by the shape of the tetrahedron with volume $v$
and with edge lengths $x_{ij}$ connecting Si$_{i}$-Si$_{j}$ nuclei.
Some edges are shorter/longer than others and they are schematically
represented by solid/dashed edges, respectively, in the middle of
Figure~\ref{fig3}. By defining an effective length as the geometric
average length of the edges $\bar{x}=(6\sqrt{2}v)^{\nicefrac{1}{3}}$
, A-D structures may be defined by simple distortion coordinates $\mathbf{Q}_{\textrm{A-D}}$
with magnitudes,

\begin{eqnarray}
Q_{\mathrm{A}} & = & +3\delta x_{12}-3\delta x_{34}\\
Q_{\mathrm{B}} & = & -\delta x_{12}+2\delta x_{13}+2\delta x_{23}-\delta x_{34}\\
Q_{\mathrm{C}} & = & +\delta x_{12}+2\delta x_{13}+2\delta x_{23}-\delta x_{34}\\
Q_{\mathrm{D}} & = & -\delta x_{12}+2\delta x_{13}-2\delta x_{23}+\delta x_{34},
\end{eqnarray}
where $\delta x_{ij}=x_{ij}-\bar{x}$ elongations have pre-factors
that depend on the number of symmetry-equivalent edges. Hence, structure
A forms a triangular pyramid with a Si$_{1}$ apex and a contracted
Si$_{2\textrm{-}4}$ base, whereas structures B, C and D form monoclinic
tetrahedrons with a $(\bar{1}010)$ mirror plane and mirror-symmetric
Si$_{3}$ and Si$_{4}$. On these three structures, we found 2, 1
and 3 contracted edges (4, 5 and 3 elongated ones), respectively.
Below, we show how these shapes are intimately related to the occupation
of the one-electron orbitals.

\begin{table}
\begin{ruledtabular}
\caption{\label{tab1}Structural details of four structure types ($R=\mathrm{A}$,
$\mathrm{B}$, $\mathrm{C}$ and $\mathrm{D}$) found for the carbon
vacancy in $4H$-SiC on different sub-lattice sites ($s=k,\,h$) and
charge states ($-2\leq q\leq+2$). Structures are specified by their
edge length variations $\delta x_{ij}$ (in Å), volume expansion $\delta v$
(in Å$^{3}$) and distortion magnitude $Q_{R}$ (in Å). Starred structures
are metastable. See text for detailed definitions.}
\begin{tabular}{llr@{\extracolsep{0pt}.}lr@{\extracolsep{0pt}.}lr@{\extracolsep{0pt}.}lr@{\extracolsep{0pt}.}lr@{\extracolsep{0pt}.}lr@{\extracolsep{0pt}.}l}
$(s,q)$ & $R$ & \multicolumn{2}{c}{$\delta x_{12}$} & \multicolumn{2}{c}{$\delta x_{13}$} & \multicolumn{2}{c}{$\delta x_{23}$} & \multicolumn{2}{c}{$\delta x_{34}$} & \multicolumn{2}{c}{$\delta v$} & \multicolumn{2}{c}{$Q_{R}$}\tabularnewline
\hline 
$(k,+\!+)$ & A & 0&032 & 0&032 & $-$0&031 & $-$0&031 & 0&626 & 0&191\tabularnewline
$(k,+)$ & A{*} & 0&061 & 0&061 & $-$0&058 & $-$0&058 & 0&140 & 0&359\tabularnewline
$(k,+)$ & B & $-$0&117 & 0&090 & 0&035 & $-$0&093 & 0&090 & 0&461\tabularnewline
$(k,0)$ & B & $-$0&241 & 0&180 & 0&139 & $-$0&200 & $-$0&443 & 1&079\tabularnewline
$(k,-)$ & C{*} & 0&000 & 0&111 & 0&112 & $-$0&309 & $-$0&679 & 0&747\tabularnewline
$(k,-)$ & D & $-$0&312 & 0&267 & $-$0&090 & 0&141 & $-$0&716 & 1&167\tabularnewline
$(k,=)$ & D & $-$0&233 & 0&345 & $-$0&234 & 0&359 & $-$1&065 & 1&750\tabularnewline
\hline 
$(h,+\!+)$ & A & 0&006 & 0&006 & $-$0&006 & $-$0&006 & 0&581 & 0&036\tabularnewline
$(h,+)$ & A & 0&061 & 0&061 & $-$0&061 & $-$0&061 & 0&196 & 0&366\tabularnewline
$(h,+)$ & B{*} & $-$0&065 & 0&088 & 0&000 & $-$0&102 & 0&132 & 0&335\tabularnewline
$(h,0)$ & B & $-$0&229 & 0&146 & 0&097 & $-$0&256 & $-$0&440 & 0&971\tabularnewline
$(h,-)$ & C & 0&037 & 0&097 & 0&068 & \multicolumn{2}{c}{$-0.367$} & $-$0&640 & 0&735\tabularnewline
$(h,-)$ & D{*} & $-$0&071 & 0&194 & -0&286 & 0&255 & $-$0&703 & 1&286\tabularnewline
$(h,=)$ & D & $-$0&220 & 0&295 & $-$0&331 & 0&292 & $-$1&033 & 1&765\tabularnewline
\end{tabular}
\end{ruledtabular}

\end{table}

In line with previous works,\cite{zywietz1999,bockstedte2003,umeda2004b,bratus2005,trinh2013}
we found that V$_{\mathrm{C}}^{+\!+}(k)$ and V$_{\mathrm{C}}^{+\!+}(h)$
defects are trigonal (structure A). Both introduce three empty states
deep in the gap, namely a singlet level below a doubly degenerate
level ($a_{1}+e$). Their separation, $\Delta_{\mathrm{xf}}$, results
from the local crystal field. Figure~\ref{fig3} represents the (hybrid)
Kohn-Sham energies of ground-state carbon vacancies as a function
of the level occupancy and sub-lattice site. The level energies are
reported with respect to the $\epsilon_{\mathrm{HOKS}}$ eigenvalue
of bulk. To make the interpretation easier, we present a spin-averaged
picture, although the filling of levels is represented with upward/downward
arrows.

From the eigenvalues we obtain crystal-field energies $\Delta_{\mathrm{xf}}(k)=0.06$~eV
and $\Delta_{\mathrm{xf}}(h)=0.30$~eV for V$_{\mathrm{C}}^{+\!+}(k)$
and V$_{\mathrm{C}}^{+\!+}(h)$, respectively. We will show that this
site-dependence of the crystal-field confers rather distinct electronic
structures on V$_{\mathrm{C}}(k)$ and V$_{\mathrm{C}}(h)$. Table~\ref{tab1}
reports the geometrical details regarding the evolution of ground-state
structures as we fill in the $a_{1}+e$ manifold with electrons. Also
included are the results for metastable structures (starred structures).
These were only found for $q=\pm1$ charge states and will be discussed
in Section~\ref{subsec:pJT}. Besides edge elongations $\delta x_{ij}$,
distortion magnitudes $Q_{R}$ and the volume expansion $\delta v=v-v_{\mathrm{bulk}}$
of the vacancy tetrahedron (with respect to the analogous quantity
in bulk) are also shown. It is clear that positive charge states are
compressive ($\delta v>0$), while negatively charged ones are tensile
($\delta v<0$). This is a consequence of the breaking/formation of
reconstructed bonds between the four Si radicals edging the vacancy
as we respectively remove/add electrons from/to defect states in the
gap. For the same reason, distortions ($Q_{R}$) tend to increase
in magnitude as we go from V$_{\mathrm{C}}^{+\!+}$ to V$_{\mathrm{C}}^{=}$.
We also note that structures A and B were concurrently found for the
positive charge state, whereas structures C and D were found for the
negative charge state. This is emphasized in Figure~\ref{fig3} by
two shaded regions.

All paramagnetic ground-states differ in their atomic geometries,
namely V$_{\mathrm{C}}^{+}(k,\mathrm{B})$, V$_{\mathrm{C}}^{+}(h,\mathrm{A})$,
V$_{\mathrm{C}}^{-}(k,\mathrm{D})$ and V$_{\mathrm{C}}^{-}(h,\mathrm{C})$,
and show $C_{1h}$, $C_{3v}$, $C_{1h}$ and $C_{1h}$ symmetry, respectively.
Inspection of the paramagnetic (highest occupied) one-electron wave-functions
allowed us to identify their symmetry and LCAO representations. For
the monoclinic structures we have two mirror-symmetric $a'$ states
(see Figure~\ref{fig3}), and they are distinguished by $-$ and
$+$ subscripts, standing for low- and high-energy symmetric states.
Hence, we found that $|a'_{-}\rangle\sim|11\bar{1}\bar{1}\rangle$
and $|a_{1}\rangle\sim|3\bar{1}\bar{1}\bar{1}\rangle$ for V$_{\mathrm{C}}^{+}(k,\mathrm{B})$
and V$_{\mathrm{C}}^{+}(h,\mathrm{A})$, respectively, while for negative
charge states we found $|a''\rangle\sim|00\bar{1}1\rangle$ and $|a'_{+}\rangle\sim|1\bar{1}00\rangle$
for V$_{\mathrm{C}}^{-}(k,\mathrm{D})$ and V$_{\mathrm{C}}^{-}(h,\mathrm{C})$,
respectively. Besides being compatible with the low-temperature EPR
and HF data, the above results reproduce earlier density-functional
findings.\cite{bockstedte2003,umeda2004b,bratus2005,trinh2013} 

We went on and explored the symmetry and wave-function character of
non-paramagnetic states. The results are shown in Figure~\ref{fig3}.
Here we can see that the evolution of the one-electron states, as
they become occupied, exhibits a rich picture, which includes crossing
and mixing (anti-crossing) features. These effects are responsible
for the structural variety that is observed, and to understand them
we have to invoke the JT and pJT effects.

\subsection{The pseudo-Jahn-Teller effect on the carbon vacancy in $4H$-SiC\label{subsec:pJT}}

While the JT theorem asserts the existence of spontaneous symmetry
breaking of degenerate electronic states, \emph{``the pJT effect is
the only source of instability and distortions of high-symmetry configurations
of polyatomic systems in non-degenerate states, and it contributes
significantly to the instability of degenerate states}''.\cite{bersuker2006} 

The pJT effect results in the softening of the adiabatic potential
energy surface (APES) around a \emph{reference configuration} $0$
with non-degenerate ground state $\Psi_{0}$, and it is due to overlap
with excited states via electron-phonon coupling. Should this softening
be severe enough to make $\Psi_{0}$ unstable against atomic distortion
towards structure $R$, the curvature of the APES along $\mathbf{Q}_{R}$,
which transforms as some irreducible representation $\Gamma_{R}$,
must be negative, $k^{R}=(\partial^{2}E/\partial\mathbf{Q}_{R}^{2})_{0}<0$.
Here $E=\langle\Psi_{0}|\hat{H}|\Psi_{0}\rangle$ is the total energy
and $\hat{H}$ the Hamiltonian. It may be shown\cite{bersuker2006,bersuker2013}
that the APES softening comes from a negative vibronic contribution
$k_{\mathrm{v}}^{R}$ to the total curvature $k^{R}=k_{0}^{R}+k_{\mathrm{v}}^{R}$,
where

\begin{equation}
k_{0}^{R}=\left\langle \Psi_{0}\left|\left(\frac{\partial^{2}\hat{H}}{\partial\mathbf{Q}_{R}^{2}}\right)_{\!\!0}\right|\Psi_{0}\right\rangle \label{eq:k0}
\end{equation}
is the harmonic curvature and from second-order perturbation theory,

\begin{equation}
k_{\mathrm{v}}^{R}=-2\sum_{n}\frac{|F_{0n}|^{2}}{E_{n}-E_{0}},\label{eq:kv}
\end{equation}
where $F_{0n}=\langle\Psi_{0}|(\partial\hat{H}/\partial\mathbf{Q}_{R})_{0}|\Psi_{n}\rangle$
are off-diagonal vibronic coupling constants between the reference
state $\Psi_{0}$ and excited states $\Psi_{n}$ with energies $E_{0}$
and $E_{n}$, respectively. $k_{0}^{R}$ represents the force constant
resisting the motion of atoms along $\mathbf{Q}_{R}$, whereas $k_{\mathrm{v}}^{R}$
is always negative and represents the change in that force constant
that results from adapting the electron distribution to one more suited
to the new nuclear coordinates,

\begin{equation}
\Psi_{R}=\Psi_{0}-\sum_{n}\frac{F_{0n}}{E_{n}-E_{0}}\Psi_{n},\label{eq:psi0prime}
\end{equation}
corresponding to a lower energy $E_{R}$. We note that unlike the
Jahn-Teller effect, the pJT effect mixes the ground state with excited
states to create new bonds and distort the structure. We may actually
state that the driving force of the pJT effect is the increase of
covalent bonding.\cite{bersuker2013}

Given that the product $(\partial\hat{H}/\partial\mathbf{Q}_{R})\,\mathbf{Q}_{R}$,
which is the linear term in the expansion of $\hat{H}$ in powers
of $\mathbf{Q}_{R}$, is fully symmetric, $\partial\hat{H}/\partial\mathbf{Q}_{R}$
must also have the same symmetry as $\mathbf{Q}_{R}$. This implies
that only excited states $\Psi_{n}$ which transform as the same irreducible
representation of $\Psi_{0}$, such that $\Gamma_{R}=\Gamma_{0}\otimes\Gamma_{n}$,
will lead to non-vanishing $F_{0n}$ coupling constants and contribute
to the softening of the APES.\cite{bader1968,pearson1969,pearson1986}

Besides the symmetry restrictions imposed to the $F_{0n}$ integral,
it is often assumed that only a few low-energy states contribute to
$k_{\mathrm{v}}^{R}$ due to the increasing $E_{n}-E_{0}$ energy
in denominator of Eq.~\ref{eq:kv}.\cite{bersuker2002} This premise
has justified the replacement of the infinite sums in Eqs.~\ref{eq:kv}
and \ref{eq:psi0prime} by a finite set of interacting states, or
indeed by a two-level paradigm where a single excited state couples
to the ground state via an effective vibronic coupling constant $k_{\mathrm{v}}^{R}=-F_{01}^{2}/\Delta$,
where $2\Delta=E_{1}-E_{0}$ is the effective energy separation between
the mixing states.\cite{bersuker2002}

For an accurate description of the pJT effect one would have to solve
the many-body Hamiltonian by accounting for dynamic correlation effects
(\emph{e.g.} by means of configuration interaction methods), the electron-phonon
coupling would have to be included as well, considering all phonons
obeying the above selection rule. Although this has been realized
for small molecules using sophisticated quantum chemistry methods,\cite{bersuker2013}
severe approximations have to be made in order to study defects in
solids. By using a single-determinant density functional approach
we may still arrive at a sufficiently detailed picture of the problem.
For instance, García-Fernández and co-workers\cite{garciafernandez2006b}
were able to explain the off-center displacement of the Fe$^{+}$
interstitial ion in SrCl$_{2}$ using local density functional theory.
On the contrary, the wave-function-based complete active space second-order
perturbation method was applied to the same problem and was unable
to reproduce the observations. This failure was attributed to the
insufficient number of states included in the active space.\cite{garciafernandez2006b}
The case of a vacancy in SiC would be much more demanding since the
active space spans many ligands to the vacancy site.

We investigated the pJT effect on the V$_{\mathrm{C}}$ defect in
$4H$-SiC, restricting our approach to a single-electron picture.
Although we do not have access to important parameters such as accurate
many-body gap energies and electron-phonon coupling strengths, we
will arrive at an instructive and reasonable picture for the observed
distortions. To that end we monitored the change of the one-electron
wave-functions and respective energies, while the atomic structure
was progressively changed from the high-symmetry V$_{\mathrm{C}}$(A)
configuration towards lower-symmetry structures B, C and D with $C_{1h}$
symmetry ($\Gamma_{R}=A'$). Therefore, according to the selection
rules, only fully symmetric states ($a'$) have to be considered as
the source of a pJT effect in V$_{\mathrm{C}}$. The calculations
reported within this Subsection were done using the spin-averaged
density-functional method within the PBE level. Some tests using a
spin-polarized HSE06 functional were also carried out, and apart from
the expected differences regarding the energy separation of levels,
the conclusions drawn below apply equally.

\begin{figure}
\includegraphics[width=8.4cm]{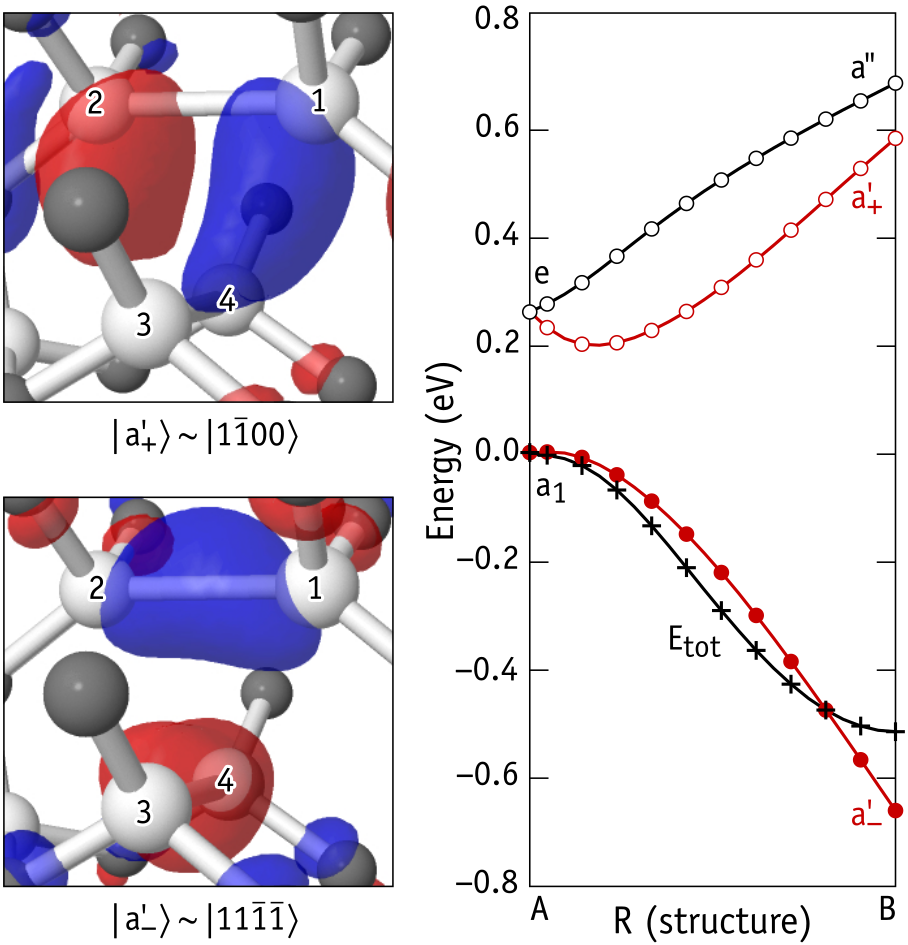}

\caption{\label{fig4}Left: shape of the highest occupied and lowest unoccupied
Kohn-Sham orbitals ($|a'_{-}\rangle$ and $|a'_{+}\rangle$, respectively)
of V$_{\mathrm{C}}^{0}(k,\mathrm{B})$ calculated at $\mathbf{k}=(0\,0\,\nicefrac{1}{2})$
within PBE-level. Blue and red isosurfaces correspond to positive
and negative phases of the orbitals. Right: Evolution of the Kohn-Sham
energies in the gap as the structure distorts from V$_{\mathrm{C}}^{0}(k,\mathrm{A})$
to the V$_{\mathrm{C}}^{0}(k,\mathrm{B})$ ground state. Occupied
and empty states are represented as solid and open circles, respectively.
The total energy ($E_{\mathrm{tot}}$) is shown as crosses. The origin
for Kohn-Sham and total energies is $\epsilon_{\mathrm{HOKS}}$ and
$E_{\mathrm{tot}}$ at $R=\mathrm{A}$, respectively. Symmetry labels
are indicated for each state.}
\end{figure}

We begin with neutral and positively charged defects. On the left
hand side of Figure~\ref{fig4} we depict the highest occupied and
lowest unoccupied Kohn-Sham states (HOKS and LUKS, respectively) for
the ground state neutral vacancy at the $k$ site, V$_{\mathrm{C}}^{0}(k,\mathrm{B})$.
Both HOKS and LUKS transform according to the $a'$ irreducible representation
of the $C_{1h}$ point group, so we differentiate them by their energy
order, \emph{i.e.} the one with lower energy is referred to as $|\textrm{HOKS}\rangle=|a'_{-}\rangle$
while the higher energy state is $|\textrm{LUKS}\rangle=|a'_{+}\rangle$.
If we consider all three states in the gap, the electronic structure
of V$_{\mathrm{C}}^{0}(k,\mathrm{B})$ is $|a_{-}'^{2}\,a'_{+}\,a''\rangle$,
where the number of electrons on a specific orbital is superscripted.
Comparing the ground state $|a'_{-}\rangle$ in Figure~\ref{fig4}
with $|a_{1}\rangle$ from V$_{\mathrm{C}}^{0}(k,\mathrm{A})$ shown
in Figure~\ref{fig1}(b), it is evident that the lower symmetry state
increases the covalent bonding between all four atoms, and that leads
to shorter Si$_{1}$-Si$_{2}$ and Si$_{3}$-Si$_{4}$ distances in
structure B. Furthermore, considering that V$_{\mathrm{C}}^{0}(k,\mathrm{A})$
is a non-degenerate ground state ($|a_{1}^{2}\,e\rangle$), we conclude
that the states exhibited in Figure~\ref{fig4} must result from
a pJT effect.

On the right hand side of Figure~\ref{fig4} we find an electronic
structure diagram, showing how the three gap states develop between
the trigonal V$_{\mathrm{C}}^{0}(k,\mathrm{A})$ state with electronic
configuration $|a_{1}^{2}\,e\rangle$ and the monoclinic V$_{\mathrm{C}}^{0}(k,\mathrm{B})$
state with electronic configuration $|a_{-}'^{2}\,a'_{+}\,a''\rangle$.
Energies of filled and empty states are represented with closed and
open symbols, respectively. The same graph also shows the total energy
change as crosses, from which we conclude that the high-symmetry configuration
A is unstable against relaxation to B. The corresponding \emph{pseudo-Jahn-Teller
relaxation energy}, $E_{\mathrm{pJT}}=0.5$~eV, relates to the added
covalence. The $\mathbf{Q}_{\mathrm{B}}$ distortion transforms as
$A'$ within $C_{1h}$ (couples to $a'$ electronic states), and consists
in the compression of Si$_{1}$-Si$_{2}$ and Si$_{3}$-Si$_{4}$
distances, along with the expansion of the remaining tetrahedron edges
(see Table~\ref{tab1}). Looking again at Figure~\ref{fig1}(b),
it becomes evident that when subject to a $\mathbf{Q}_{\mathrm{B}}$
distortion, the state $|a_{1}\rangle\sim|3\bar{1}\bar{1}\bar{1}\rangle$,
which transforms as $a'$ within $C_{1h}$ and shows a strong anti-bonding
character between Si$_{1}$-Si$_{2}$, should raise in energy, while
the doublet component $|e'\rangle\sim|0\bar{2}11\rangle$, also transforming
as $a'$ within $C_{1h}$ and showing a bonding character between
Si$_{3}$-Si$_{4}$, is expected to be stabilized and lower its energy.
This opposite coupling leads to the typical pJT \emph{anti-crossing}
pattern shown in Figure~\ref{fig4} for $|a'_{-}\rangle$ and $|a'_{+}\rangle$
states.

We may estimate the relative contribution (mixing) from $|a_{1}\rangle$
and $|e'\rangle$ states to the pJT distorted $|a'_{-}\rangle$ and
$|a'_{+}\rangle$ states using our simple LCAO model. From inspection
of Figures~\ref{fig1}(b) and \ref{fig4}, and considering normalization
coefficients $|a_{1}\rangle=12{}^{-\nicefrac{1}{2}}\,|3\bar{1}\bar{1}\bar{1}\rangle$
and $|e'\rangle=6^{-\nicefrac{1}{2}}\,|0\bar{2}11\rangle$ we arrive
at,

\begin{eqnarray}
|a'_{+}\rangle\! & =\!(2/3)^{\nicefrac{1}{2}}|a_{1}\rangle+(1/3)^{\nicefrac{1}{2}}|e'\rangle=\! & (1/2)^{-\nicefrac{1}{2}}\,|1\bar{1}00\rangle\label{eq:a+}\\
|a'_{-}\rangle\! & =\!(1/3)^{\nicefrac{1}{2}}|a_{1}\rangle-(2/3)^{\nicefrac{1}{2}}|e'\rangle=\! & 1/2\,|11\bar{1}\bar{1}\rangle.\label{eq:a-}
\end{eqnarray}

Like the isosurfaces shown in Figure~\ref{fig4}, the ground state
$|a'_{-}\rangle$ in Eq.~\ref{eq:a-} has the same phase (bonding
character) on Si$_{1,2}$ and Si$_{3,4}$ atom pairs, and that mostly
comes from $|e'\rangle$. Conversely, $|a'_{+}\rangle$ is an anti-bonding
state between Si$_{1,2}$ atoms with vanishing amplitude on Si$_{3,4}$,
and most of its character comes from $|a_{1}\rangle$. The above discussion
and conclusions can be applied to the neutral vacancy at the hexagonal
site as well. However, the stronger crystal-field splitting leads
to a larger energy gap $2\Delta=E_{1}-E_{0}$, and therefore to a
weaker mixing effect.

\begin{figure}
\includegraphics[width=8.4cm]{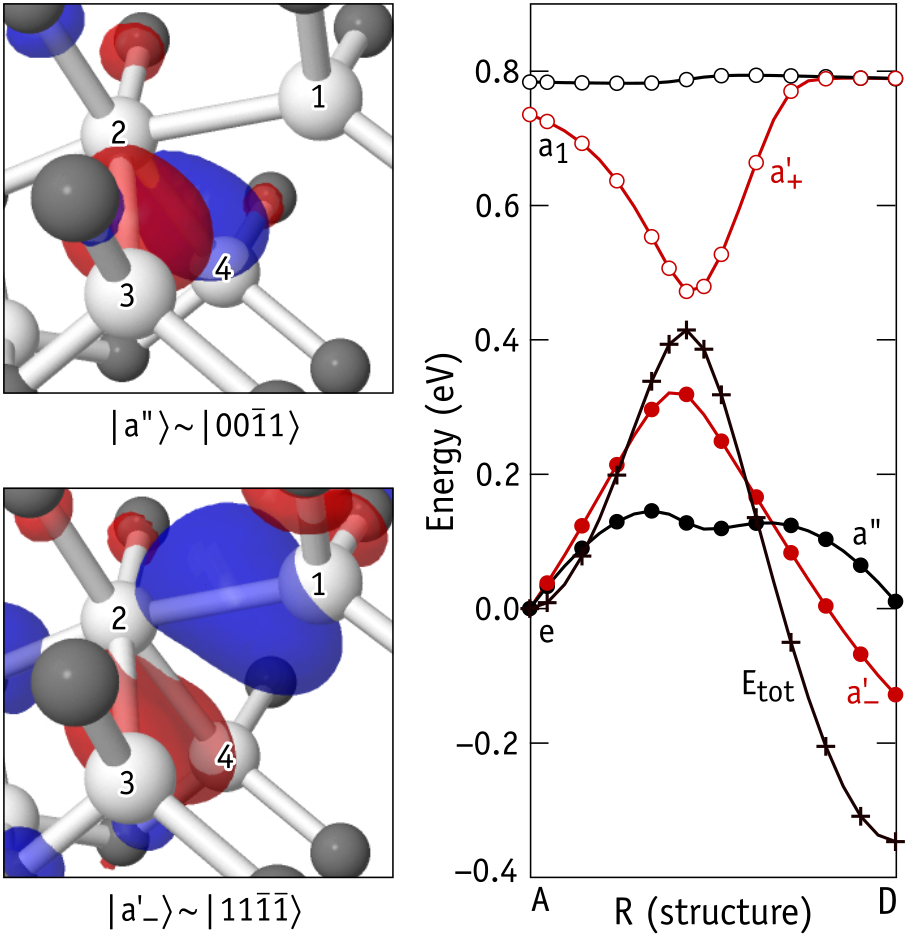}

\caption{\label{fig5}Left: shape of the HOKS$-1$ and HOKS orbitals ($|a'_{-}\rangle$
and $|a''\rangle$, respectively) of V$_{\mathrm{C}}^{=}(k,\mathrm{D})$
calculated at $\mathbf{k}=(0\,0\,\nicefrac{1}{2})$ within PBE-level.
Blue and red isosurfaces correspond to positive and negative phases
of the orbitals. Right: Evolution of the Kohn-Sham energies in the
gap as the structure distorts from V$_{\mathrm{C}}^{=}(k,\mathrm{A})$
to the V$_{\mathrm{C}}^{=}(k,\mathrm{D})$ ground state. Occupied
and empty states are represented as solid and open circles, respectively.
The total energy ($E_{\mathrm{tot}}$) is shown as crosses. The origin
for Kohn-Sham and total energies is $\epsilon_{\mathrm{HOKS}}$ and
$E_{\mathrm{tot}}$ at $R=\mathrm{A}$, respectively. The topmost
data points connected by a flat curve represent the conduction band
bottom. Symmetry labels are indicated for each state.}
\end{figure}

For positively charged vacancies on both $k$- and $h$-sites, the
shape of the electronic structure diagrams (and wave-functions) were
found to be close to those of Figure~\ref{fig4}, although $E_{\mathrm{tot}}$
for V$_{\mathrm{C}}^{+}$(A) and V$_{\mathrm{C}}^{+}$(B) indicated
that these were both minima in the APES of $k$ and $h$ sites. From
Table~\ref{tab1} we see that the distortion magnitudes of positively
charged defects are considerably smaller than in neutral defects.
We may conclude that the pJT coupling is weaker for V$_{\mathrm{C}}^{+}$(B),
particularly in the hexagonal site where the crystal field is stronger.
Here, the V$_{\mathrm{C}}^{+}(h,\mathrm{B})$ state $|a_{-}'^{1}\rangle\sim|11\bar{1}\bar{1}\rangle$
with two (weak) Si-Si bonds sharing a single electron, is essentially
degenerate with the V$_{\mathrm{C}}^{+}(h,\mathrm{A})$ state $|a_{1}^{1}\rangle\sim|3\bar{1}\bar{1}\bar{1}\rangle$.
Their energy difference is estimated below $E_{\mathrm{pJT}}=1$~meV.

For the negatively charged vacancies, the picture is dramatically
different. We start by analyzing the double negative charge state,
where structure D was found to be the most stable for both $k$ and
$h$ sites. For the trigonal structure on the $k$-site (the structure
was relaxed by symmetrizing the forces), we found that the $|a_{1}^{2}e^{2}\rangle$
state with spin-0 was less stable than the spin-1 configuration by
0.20~eV, but the latter was still metastable by 0.15~eV when compared
to the $|e^{4}a_{1}\rangle$ non-degenerate ground state. On the left
hand side of Figure~\ref{fig5} we depict the (fully occupied) levels
found within the gap for V$_{\mathrm{C}}^{=}(k,\mathrm{D})$. Comparing
these wave-functions with those shown in Figure~\ref{fig1}(b) for
the symmetric structure, we realize that although the HOKS state $|a''\rangle$
of structure D is rather similar to $|e''\rangle\sim|00\bar{1}1\rangle$
from the trigonal structure, the $|a'_{-}\rangle$ does not find a
good match, although one could suggest some resemblance with $|a_{1}\rangle$.
Considering that (i) V$_{\mathrm{C}}^{=}(k,\mathrm{A})$ is non-degenerate,
and therefore not vulnerable to a JT distortion, and (ii) that $|a'_{-}\rangle$
is a mixed state with a major contribution from $|a_{1}\rangle$,
the wave-functions exhibited in Figure~\ref{fig5} must result from
a pJT effect. In fact, looking at the right hand side of Figure~\ref{fig5},
it becomes evident that $|a_{1}\rangle$ has been converted into $|a'_{-}\rangle$
under distortion $\mathbf{Q}_{\mathrm{D}}$, whereas $|a'_{+}\rangle$
(derived from $e'$) seems to have merged into the conduction band
(uppermost state close to 0.8~eV) before the ground state was attained.

In fact, $|a'_{-}\rangle$ shows bonding character for Si$_{1}$-Si$_{2}$
and Si$_{3}$-Si$_{4}$ pairs, and can be described approximately
as $\sim|11\bar{1}\bar{1}\rangle$ like in Eq.~\ref{eq:a-}. It becomes
now clear that structure D results from structure B (occupation of
$|a'_{-}\rangle$ leads to the shortening of Si$_{1}$-Si$_{2}$ and
Si$_{3}$-Si$_{4}$ distances) combined with the occupation of $|a''\rangle$,
which is anti-bonding on Si$_{3}$-Si$_{4}$. The result is a tetrahedral
structure with short Si$_{1}$-Si$_{2}$, Si$_{2}$-Si$_{3}$ and
Si$_{2}$-Si$_{4}$ edges. We finally note that V$_{\mathrm{C}}^{=}(h)$
shows a similar behavior to V$_{\mathrm{C}}^{=}(k)$, with the metastability
of the trigonal structure by 0.5~eV being worthy of mentioning.

\begin{figure}
\includegraphics[width=8.4cm]{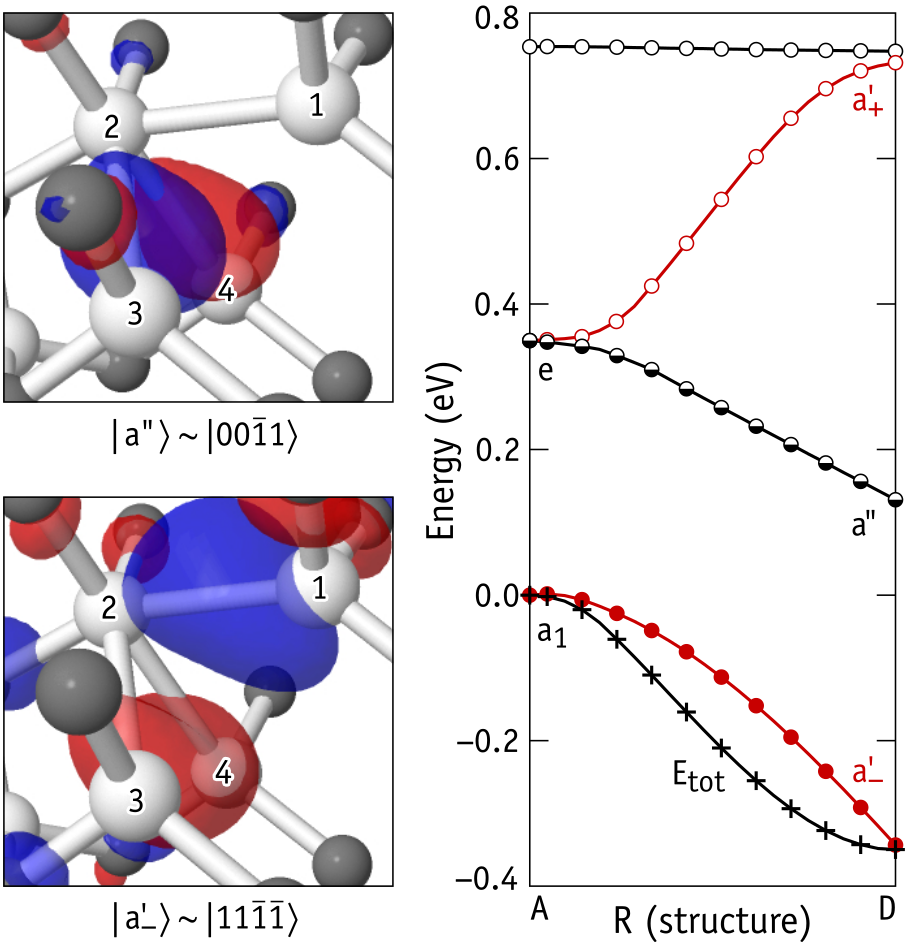}

\caption{\label{fig6}Left: shape of the HOKS$-1$ and HOKS orbitals ($|a'_{-}\rangle$
and $|a''\rangle$, respectively) of V$_{\mathrm{C}}^{-}(k,\mathrm{D})$
calculated at $\mathbf{k}=(0\,0\,\nicefrac{1}{2})$ within PBE-level.
Blue and red isosurfaces correspond to positive and negative phases
of the orbitals. Right: Evolution of the Kohn-Sham energies in the
gap as the structure distorts from V$_{\mathrm{C}}^{-}(k,\mathrm{A})$
to the V$_{\mathrm{C}}^{-}(k,\mathrm{D})$ ground state. Occupied,
semi-occupied and empty states are represented as solid, half-filled
and open circles, respectively. The total energy ($E_{\mathrm{tot}}$)
is shown as crosses. The origin for Kohn-Sham and total energies is
$\epsilon_{\mathrm{HOKS}}$ and $E_{\mathrm{tot}}$ at $R=\mathrm{A}$,
respectively. The topmost data points connected by a flat curve represent
the conduction band bottom. Symmetry labels are indicated for each
state.}
\end{figure}

For the singly negative charge states, we found that imposing structure
A to the defect (symmetry-constrained relaxation) the self-consistent
electronic structure showed a $|a_{1}^{2}\,e^{1}\rangle$ occupation
(with spin-1/2) for both sites $h$ and $k$. In this case both trigonal
structures are vulnerable to the JT effect. Monoclinic distortions
applied to V$_{\mathrm{C}}^{-}(k,\mathrm{D})$ and V$_{\mathrm{C}}^{-}(h,\mathrm{C})$
ground states were found to release 0.39~eV and 0.36~eV, respectively.
The two highest occupied states of V$_{\mathrm{C}}^{-}(k,\mathrm{D})$
and their change along the A-D path on the APES are depicted in Figure~\ref{fig6}.
We can conclude that despite showing occupied $|a'_{-}\rangle$ and
$|a''\rangle$ states like in the double minus charge state (and hence
showing a similar structure), the diagram on the right side of the
figure is rather different than that shown in Figure~\ref{fig5}.
The high-symmetry configuration V$_{\mathrm{C}}^{-}(k,\mathrm{A})$
is now unstable due to the Jahn-Teller effect. Interestingly, the
splitting order of the $e$-state favors the stabilization of the
nodal $a''$ state with higher kinetic-energy (under the monoclinic
field). We will come back to this issue in Section~\ref{sec:discussion}.

On the $h$ site we found that the JT splitting order of the $e$-state
of V$_{\mathrm{C}}^{-}(h,\mathrm{A})$ involves the raising in energy
of the anti-symmetric state $a''$ under the monoclinic distortion
$\mathbf{Q}_{\mathrm{C}}$. Unlike for the cubic site, we have now
a $|a_{-}'^{2}\,a_{+}'^{1}\,a''\rangle$ occupation scheme. This difference
is attributed to the relatively stronger crystal field separating
$a_{1}$ and $e$ states of the symmetric configuration in V$_{\mathrm{C}}^{-}(h)$,
and consequently to a weaker coupling between $|a'_{-}\rangle$ and
$|a'_{+}\rangle$ states. The resulting structure C is therefore based
on structure B (due to the occupation of $|a'_{-}\rangle$), but shows
an elongated Si$_{1}$-Si$_{2}$ distance due to occupation of the
anti-bonding $|a'_{+}\rangle$ state (see Figure~\ref{fig4}).

\subsection{Dynamical effects}

We calculated minimum energy barriers separating different distorted
structures along the APES using the NEB method. Table~\ref{tab2}
reports the most favorable forward ($E_{\mathrm{fwd}}$) and backward
($E_{\mathrm{bak}}$) barriers, between several initial and final
structures. The sub-lattice site and charge state are indicated as
$(s,q)$ pairs on the first column. Two types of mechanisms were considered,
namely rotations (R) and transformations (T). A rotation involves
a 120$^{\circ}$ \emph{rotation} of the mirror plane of monoclinic
structures (B, C and D), which are converted into symmetry-equivalent
final states (B', C' and D'). A transformation involves a structural
change to an inequivalent state ($R_{\mathrm{i}}\neq R_{\mathrm{f}}$),
which may as well include a change in the direction of the symmetry
plane. In that case, they are also indicated by \emph{primed} final
states.

\begin{table}
\caption{\label{tab2}Forward ($E_{\mathrm{fwd}}$) and backward ($E_{\mathrm{bak}}$)
transition barriers between symmetry equivalent (Type R - rotation)
and inequivalent (Type T - transformation) states of the V$_{\mathrm{C}}$
defect in $4H$-SiC. The sub-lattice site and charge state are shown
on the first column. $E_{\mathrm{i}}$ and $E_{\mathrm{f}}$ are initial
and final energies, respectively. In Type-R mechanisms, the final
state $V_{\mathrm{C}}^{q}(s,R_{\mathrm{f}})$ has higher energy than
the initial state $V_{\mathrm{C}}^{q}(s,R_{\mathrm{i}})$. Primed
$R_{\mathrm{f}}$ structures indicate a change in the orientation
of the mirror plane. All data are in eV.}
\begin{ruledtabular}
\begin{tabular}{ccccccc}
$(s,q)$ & Type & $R_{\mathrm{i}}$ & $R_{\mathrm{f}}$ & $E_{\mathrm{fwd}}$ & $E_{\mathrm{bak}}$ & $E_{\mathrm{f}}-E_{\mathrm{i}}$\tabularnewline
\hline 
$(k,+)$ & R & $\mathrm{B}$ & $\mathrm{B}'$ & 0.05 &  & \tabularnewline
$(k,+)$ & T & $\mathrm{B}$ & $\mathrm{A}$ & 0.12 & 0.04 & 0.08\tabularnewline
$(k,0)$ & R & $\mathrm{B}$ & $\mathrm{B}'$ & 0.41 &  & \tabularnewline
$(k,-)$ & R & $\mathrm{C}$ & $\mathrm{C}'$ & 0.15 &  & \tabularnewline
$(k,-)$ & R & $\mathrm{D}$ & $\mathrm{D}'$ & 0.17 &  & \tabularnewline
$(k,-)$ & T & $\mathrm{D}$ & $\mathrm{C}$ & 0.14 & 0.12 & 0.02\tabularnewline
$(k,-)$ & T & $\mathrm{D}$ & $\mathrm{C}'$ & 0.06 & 0.04 & 0.02\tabularnewline
$(k,=)$ & R & $\mathrm{D}$ & $\mathrm{D}'$ & 0.28 &  & \tabularnewline
\hline 
$(h,+)$ & R & $\mathrm{B}$ & $\mathrm{B}'$ & 0.02 &  & \tabularnewline
$(h,+)$ & T & $\mathrm{A}$ & $\mathrm{B}$ & 0.02 & 0.02 & $<$0.01\tabularnewline
$(h,0)$ & R & $\mathrm{B}$ & $\mathrm{B}'$ & 0.30 &  & \tabularnewline
$(h,-)$ & R & $\mathrm{C}$ & $\mathrm{C}'$ & 0.08 &  & \tabularnewline
$(h,-)$ & R & $\mathrm{D}$ & $\mathrm{D}'$ & 0.05 &  & \tabularnewline
$(h,-)$ & T & $\mathrm{C}$ & $\mathrm{D}$ & 0.24 & 0.10 & 0.14\tabularnewline
$(h,-)$ & T & $\mathrm{C}$ & $\mathrm{D}'$ & 0.17 & 0.03 & 0.14\tabularnewline
$(h,=)$ & R & $\mathrm{D}$ & $\mathrm{D}'$ & 0.25 &  & \tabularnewline
\end{tabular}
\end{ruledtabular}

\end{table}

For V$_{\mathrm{C}}^{+}(k)$ we have two low energy structures, namely
A (metastable) and B (ground state). The simple $\mathrm{B}\leftrightarrow\mathrm{B}'$
rotation mechanism involves surmounting a small 0.05~eV barrier.
On the other hand, the $\mathrm{B}\rightarrow\mathrm{A}$ transformation
has a 0.12~eV barrier, and so it has the alternative $\mathrm{B}\rightarrow\mathrm{A}\rightarrow\mathrm{B}'$
combined rotation mechanism. For V$_{\mathrm{C}}^{+}(h)$, structure
A was found to be more stable than B by less than 1~meV, so we consider
them essentially degenerate. Both rotation of the mirror plane in
$\mathrm{B}\leftrightarrow\mathrm{B}'$ as well as the $\mathrm{A}\rightarrow\mathrm{B}$
transformation involve overcoming a minute barrier of 0.02~eV. Hence,
for the $k$-site, the rotation between (symmetric) equivalent B structures
should be the first dynamic effect to take place as the temperature
is raised from 5~K. On the other hand, for the $h$-site it appears
that even at very low temperatures, V$_{\mathrm{C}}^{+}(h,\mathrm{A})$
defects may cohabit with V$_{\mathrm{C}}^{+}(h,\mathrm{B})$ states,
with the later being able to hop between different alignments.

V$_{\mathrm{C}}^{-}(k)$ was found to have low energy in structures
C (metastable) and D (ground state), which are separated by only 0.02~eV.
Simple rotation mechanisms $\mathrm{D}\leftrightarrow\mathrm{D}'$
and $\mathrm{C}\leftrightarrow\mathrm{C}'$ involve barriers of 0.17~eV
and 0.15~eV, respectively. The in-plane $\mathrm{D}\rightarrow\mathrm{C}$
transformation also has a comparable barrier of 0.14~eV. On the other
hand, the off-plane transformation $\mathrm{D}\rightarrow\mathrm{C}'$
is the most favorable mechanism with a barrier of 0.06~eV. These
results suggest that the lowest-temperature dynamic mechanism involving
atomic motion in V$_{\mathrm{C}}^{-}(k)$ should involve a $\mathrm{D}\rightarrow\mathrm{C}'\rightarrow\mathrm{D}''\rightarrow\cdots$
sequential transformation. For the $h$-site, the negatively charged
vacancy is also stable for structures C and D, although D is now metastable
by 0.14~eV. For this reason, in-plane $\mathrm{C}\rightarrow\mathrm{D}$
and off-plane $\mathrm{C}\rightarrow\mathrm{D}'$ transformations
involve relatively high barriers of 0.24~eV and 0.17~eV, respectively,
whereas simple rotation mechanisms $\mathrm{C}\leftrightarrow\mathrm{C}'$
and $\mathrm{D}\leftrightarrow\mathrm{D}'$ have only to overcome
0.08~eV and 0.05~eV barriers. This suggests that at low temperatures,
the first thermally-activated dynamic effect will involve a simple
$\mathrm{C}\leftrightarrow\mathrm{C}'$ rotations.

We note that several of the above figures, like $\mathrm{V}{}_{\mathrm{C}}^{-}(k,\mathrm{D})\rightarrow\mathrm{V}{}_{\mathrm{C}}^{-}(k,\mathrm{C}')$
or $\mathrm{V}{}_{\mathrm{C}}^{-}(h,\mathrm{D})\rightarrow\mathrm{V}{}_{\mathrm{C}}^{-}(h,\mathrm{D}')$
barriers are rather small. They are close to the error bar of the
current methodology and should be considered with caution. However,
their relative magnitudes are in line with the lowest-temperature
dynamic processes observed in the EPR main signals and hyperfines.
Accordingly, raising the temperature above 50~K, the pattern of the
V$_{\mathrm{C}}^{+}(k)$ main line is converted from monoclinic to
trigonal. This is assigned to a $\mathrm{B}\leftrightarrow\mathrm{B}'$
rotation with a calculated 0.05~eV barrier (estimated experimentally
as 14~meV). Above 10~K, the main EPR signal of V$_{\mathrm{C}}^{+}(h)$
and related HFs suffer a progressive change. Such low temperature
is consistent with the minute (0.02~eV) $\mathrm{A}\rightarrow\mathrm{B}$
transformation barrier. Raising the temperature above 40~K, the V$_{\mathrm{C}}^{-}(k)$
signal shows a series of different transformations, which can be explained
by a sequence $\mathrm{D}\rightarrow\mathrm{C}'\rightarrow\mathrm{D}''\cdots$
of transformations with a 0.06~eV barrier. Finally, for V$_{\mathrm{C}}^{-}(h)$,
the measurements indicate that the first thermally activated process
is limited by an estimated barrier of 20~meV at about 60-70~K, also
in line with our calculated barrier of 0.08~eV for the $\mathrm{C}\leftrightarrow\mathrm{C}'$
realignment.

\begin{figure}
\includegraphics[width=8.5cm]{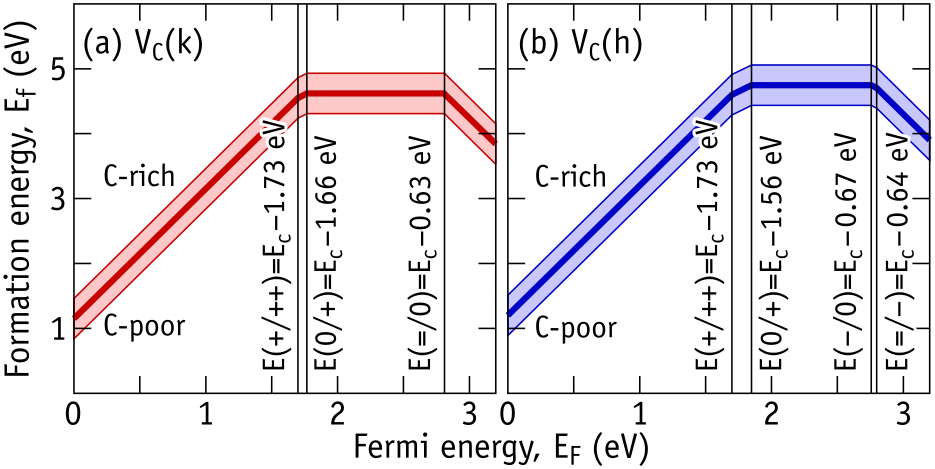}

\caption{\label{fig7}Formation energy ($E_{\mathrm{f}}$) of the carbon vacancy
at the cubic (a) and hexagonal (b) sites as a function of the Fermi
energy ($E_{\mathrm{F}}$). Lower, central and upper lines represent
$E_{\mathrm{f}}$ values for crystals grown under C-poor (or Si-rich),
stoichiometric and C-rich conditions.}
\end{figure}

The neutral charge states (both at $k$ and $h$ sites) only have
one stable structure and only $\mathrm{B}\leftrightarrow\mathrm{B}'$
rotations are possible. For these mechanisms we found relatively high
barriers of about 0.4~eV and 0.3~eV for V$_{\mathrm{C}}^{0}(k)$
and V$_{\mathrm{C}}^{0}(h)$, respectively. For double negatively
charged defects we also found relatively static defects. Here the
ground state is the D structure for V$_{\mathrm{C}}^{=}(k)$ and V$_{\mathrm{C}}^{=}(h)$,
with respective metastable C structures at 0.26~eV and 0.22~eV above
D. Their respective $\mathrm{D}\leftrightarrow\mathrm{D}'$ rotation
mechanism were found to be limited by 0.28 eV and 0.25~eV high barriers.

\subsection{Electrical levels and metastability}

The formation energy of V$_{\mathrm{C}}$ defects was calculated using
Eq.~\ref{eq:formation}. The results are depicted in Figure~\ref{fig7},
where each diagram includes formation energies under C-rich, stoichiometric
and C-poor growth-conditions. In agreement with Ref.~\onlinecite{hornos2011},
the formation energy of neutral V$_{\mathrm{C}}^{0}(k)$ and V$_{\mathrm{C}}^{0}(h)$
defects in C-rich material is 4.93~eV and 5.06~eV, respectively,
whereas in C-poor $4H$-SiC these quantities are off-set by $\Delta H_{\mathrm{SiC}}^{\mathrm{f}}=-0.62$~eV
to 4.31~eV and 4.44~eV, respectively. For stoichiometric conditions
$E_{\mathrm{f}}$ values are mid-way between C-rich and C-poor figures.
The C-rich results agree very well with the formation enthalpy of
4.8-5.0~eV measured from samples grown under analogous conditions.\cite{ayedh2014,ayedh2015}

\begin{figure*}
\includegraphics[width=18cm]{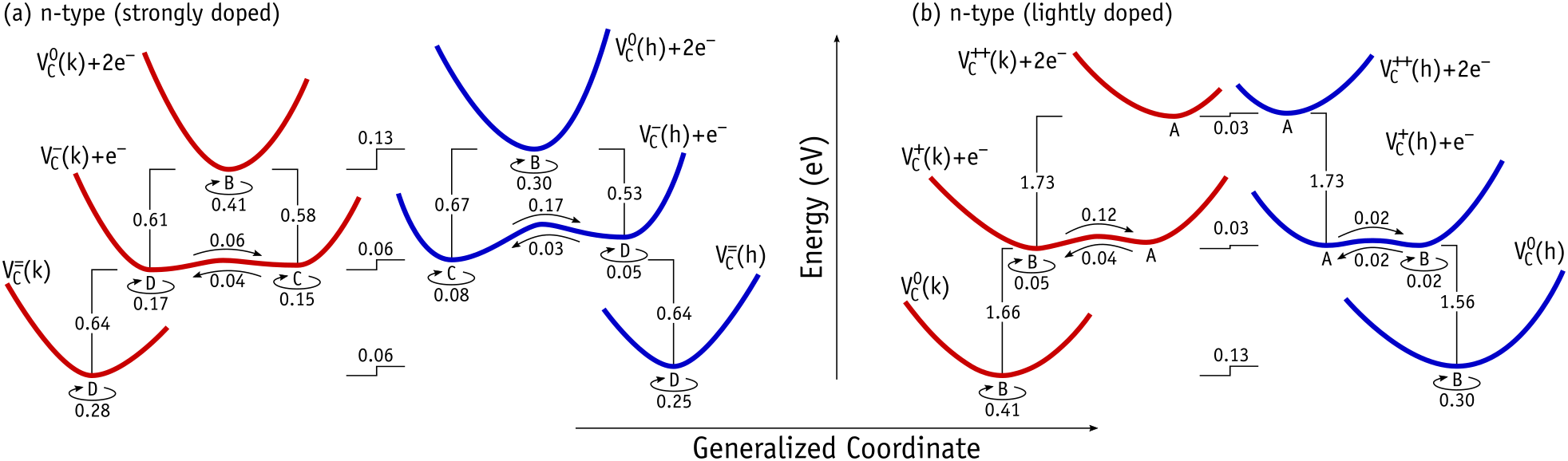}

\caption{\label{fig8}Configuration coordinate diagram of V$_{\mathrm{C}}$
defects in $4H$-SiC. Acceptor and donor transitions are shown in
(a) and (b), representing strongly doped and lightly doped n-type
material, respectively. Electronic transitions (up in energy) involve
the emission of one electron to the conduction band ($e^{-}$). Insets
(a) and (b) include two diagrams, one for each sub-lattice site. Energies
(in eV) are accompanied with guidelines for better perception. Forward/backward
transformation and rotation barrier energies are indicated by right/left
and spinning arrows, respectively. Step-like guidelines indicate the
energy difference between ground states of defects in $k$ and $h$
sites. Minima of the potential curves represent stable/metastable
structures and are identified with labels A-D.}
\end{figure*}

Also like in Ref.~\onlinecite{hornos2011}, we only find negative-$U$
behavior ($U=-0.03$~eV) for the acceptor levels of V$_{\mathrm{C}}(k)$,
\emph{i.e.} we find a $E(=/0)=E_{\mathrm{c}}-0.63$~eV occupancy
level, with V$_{\mathrm{C}}^{-}(k)$ being metastable irrespectively
of the position of the Fermi energy. The acceptors of V$_{\mathrm{C}}(h)$
are located close to $E_{\mathrm{c}}-0.6$~eV and separated by a
rather small but positive $U=0.03$~eV. Donor levels are estimated
between $E_{\mathrm{c}}-1.56$~eV and $E_{\mathrm{c}}-1.73$~eV,
with $E(0/+)$ and $E(+/+\!+)$ levels being separated by positive
$U=0.07$~eV and $0.17$~eV for V$_{\mathrm{C}}(k)$ and V$_{\mathrm{C}}(h)$
defects, respectively.

We may conclude that both acceptor and donor levels are very close
to the DLTS measurements of Z$_{1/2}$ at $E_{\mathrm{c}}-0.5\textrm{-}0.7$~eV
and EH$_{6/7}$ at about $E_{\mathrm{c}}-1.4\textrm{-}1.5$~eV, respectively,
supporting the assignment of both signals to the carbon vacancy. The
relative magnitude of the calculated $U$-values can be connected
with the amount of excitation, namely the illumination frequency/intensity
and sample temperature needed for the observation of the paramagnetic
states. Accordingly, V$_{\mathrm{C}}^{-}(k)$ is predicted to have
$U<0$~eV and its EPR signal could only be seen in highly-doped n-type
material under illumination,\cite{son2012} traces of the V$_{\mathrm{C}}^{-}(h)$
signal, with calculated $U\approx0$~eV, were detected in darkness
at $T>90$~K,\cite{son2012} both EPR signals of V$_{\mathrm{C}}^{+}$,
with calculated $U>0$~eV, could be detected in darkness even at
temperatures as low as $T=5$~K.\cite{umeda2004b} This trend agrees
with our calculated ordering of $U$ values, so that paramagnetic
states with smaller (and negative) $U$ values have lower probability
to occur because of concurrent formation of energetically favorable
diamagnetic states.

While singly negative charge states need some sort of excitation (optical
or thermal) to be observed, which is consistent with a negative-$U$
behavior, the EPR data for positively charged vacancies seems actually
characteristic of a positive-$U$ ordering of levels. Otherwise, how
could we explain the observation of both V$_{\mathrm{C}}^{+}(k)$
and V$_{\mathrm{C}}^{+}(h)$ at $T=5$~K without illumination?\cite{umeda2004b}
To investigate this issue we need to have a more detailed view of
the electronic/atomic transitions that take place during DLTS measurements.
Combining the relative energies of V$_{\mathrm{C}}^{q}(s,R)$ states
with the calculated rotation/transformation barriers and levels, we
arrived at the configuration coordinate diagram of Figure~\ref{fig8},
which describes several electronic emission processes that take place
during electrical measurements in n-type material.

Figure~\ref{fig8}(a) represents electron emission energies from
negative charge states in highly doped n-type material. Here, the
Fermi level is located above the acceptor levels. Figure~\ref{fig8}(b)
represents electron emission energies from donor states in lightly
doped n-type material. The Fermi level is now between donor and acceptor
levels. Energies on each inset {[}(a) and (b){]} refer to different
scales, and the curvature of the potential curves is arbitrary.

We start by analyzing electron emissions in highly doped material
from double negatively charge vacancies on the $k$-site. In a DLTS
measurement, under a zero-bias filling pulse, all V$_{\mathrm{C}}(k)$
electron traps will be filled with electrons and vacancies will be
found in the V$_{\mathrm{C}}^{=}(k,\mathrm{D})$ state. Around room
temperature under reverse-bias,\cite{hemmingsson1997} after a first
electron emission with binding energy calculated as 0.64~eV, a second
emission with lower energy (0.61~eV) will follow immediately, and
the defect will end up in the neutral charge state. The reason for
the negative-$U$ behavior of the V$_{\mathrm{C}}(k)$ acceptor sequence
is the strong relaxation to structure B after the second emission.

On the hexagonal site, the electron binding energy of V$_{\mathrm{C}}^{=}(h)$
is 0.64~eV considering a transition between ground states, $\mathrm{V}{}_{\mathrm{C}}^{=}(h,\mathrm{D})\rightarrow\mathrm{V}{}_{\mathrm{C}}^{-}(h,\mathrm{C})+e^{-}$
. The second emission from V$_{\mathrm{C}}^{-}(h,\mathrm{C})$ has
a slightly larger binding energy of 0.67~eV, corresponding to a small
but positive-$U$ previously shown in Figure~\ref{fig7}(b). This
discrepancy with the DLTS data could indicate that the stability of
V$_{\mathrm{C}}^{-}(h,\mathrm{C})$ is overestimated, or alternatively,
the stability of V$_{\mathrm{C}}^{0}(h,B)$ is underestimated.

It is possible that the conversion between C and D configurations
could play a role during electron emission. For instance, the calculated
energy barrier between V$_{\mathrm{C}}^{-}(k,\mathrm{D})$ and V$_{\mathrm{C}}^{-}(k,\mathrm{C})$
is 0.06~eV, which should be compared with 0.02~eV from analysis
of the thermally activated motional effects observed in EPR.\cite{umeda2005}
Since atomic rotations/transformations occur on a much faster time-scale
than electronic transitions, we cannot exclude the possibility that
at the temperature of the DLTS measurements, emission from V$_{\mathrm{C}}^{-}$
could initiate from metastable states, thus leading to an effective
smaller second ionization.

In as-grown material, where vacancies are in equilibrium conditions,
Ref.~\onlinecite{hemmingsson1997} reports a Z$_{2}(-/0)$ peak which
was about twice the intensity of Z$_{1}(-/0)$. This contrasts with
irradiated material where Z$_{1}(-/0)$ and Z$_{2}(-/0)$ show up
with about the same magnitude. Since carbon vacancies are invariably
more stable in the cubic site (by about 0.06~eV in n-type SiC as
shown in Figure~\ref{fig8}), we attribute Z$_{2}$ and Z$_{1}$
signals to V$_{\mathrm{C}}(k)$ and V$_{\mathrm{C}}(h)$, respectively.
A 1:2 intensity ratio corresponds to a concentration ratio $[\mathrm{V}{}_{\mathrm{C}}(h)]:[\mathrm{V}{}_{\mathrm{C}}(k)]=\exp(-0.06/k_{\mathrm{B}}T)$
under equilibrium conditions at $T\approx1000$~K. This assignment
is also consistent with a previous connection between Z$_{1}$ and
V$_{\mathrm{C}}(h)$ by photo-EPR,\cite{son2012} it is supported
by the lower calculated $U$-value for the acceptor levels of V$_{\mathrm{C}}(k)$
{[}when compared to V$_{\mathrm{C}}(h)${]}, and it agrees with the
calculated deeper $(-/0)$ transition for V$_{\mathrm{C}}(h)$ than
V$_{\mathrm{C}}(k)$. We finally note that the calculated $U=+0.03$~eV
between $(-/0)$ and $(=/-)$ acceptors of V$_{\mathrm{C}}(h)$ must
not be far from the true value. This value is consistent with the
observation of the V$_{\mathrm{C}}^{-}(h)$ EPR signal at 100~K (without
illumination) and its photo-ionization with photon energies $h\nu>0.74$~eV
below the band-gap threshold.\cite{son2012} In contrast, the analogous
photo-ionization for the cubic vacancy could not be observed, most
probably because $U$ is more negative and the most favorable V$_{\mathrm{C}}(k)$
defects are diamagnetic.

The calculated donor levels are also represented in Figure~\ref{fig8}(b).
Although first and second electron binding energies, i.e. $(0/+)$
and $(+/+\!+)$ transitions, are close for both sub-lattice sites,
they do not form a negative-$U$ sequence. The $U$-value for $V_{\mathrm{C}}(k)$
is $1.73-1.66=0.07$~eV, while for the $h$-site $U=0.17$~eV. These
results are in partial agreement with the data reported by Booker
\emph{et~al.}\cite{booker2016}, where $U=-0.04$~eV for EH7 and
$U=0.03\pm0.04$~eV for EH6, obviously favoring the assignment of
V$_{\mathrm{C}}(k)$ and V$_{\mathrm{C}}(h)$ to EH7 and EH6, respectively.
Again, we may use the relative stability of $\mathrm{V}_{\mathrm{C}}(k)$
and $\mathrm{V}_{\mathrm{C}}(h)$ to identify the sub-lattice sites
of EH7 and EH6 signals. In as-grown material and electron irradiated
samples subject to high temperature anneals, the EH6:EH7 ratio was
found to be about 4:5,\cite{danno2006} suggesting that EH7 is more
stable. The relative energies in Figure~\ref{fig8}(b), confirm that
EH$_{6}$ and EH$_{7}$ should therefore be connected toV$_{\mathrm{C}}(h)$
and V$_{\mathrm{C}}(k)$, respectively.

Of course there is some degree of uncertainty in the calculated magnitudes
(and sign) of the $U$-values. Within the present level of theory,
electronic levels are usually affected by error bars of about $\sim0.1$~eV
due to spurious (strain, Coulomb or dispersive) periodic interactions.
However, as we will point out in the next Section, there is further
experimental evidence for a positive-$U$ ordering of donor transitions.

\section{Discussion\label{sec:discussion}}

We start by addressing the strong temperature dependence of the EPR
data related to V$_{\mathrm{C}}^{+}(h)$. The V$_{\mathrm{C}}^{+}(h)$
defect in $4H$-SiC has the properties of a pJT distorted structure
in the weak-coupling regime, where the pseudo-Jahn-Teller relaxation
energy is much smaller than the Debye frequency, $E_{\mathrm{pJT}}\ll\hbar\omega_{\mathrm{D}}$.\cite{bersuker2006}
For the case of $4H$-SiC we have $\hbar\omega_{\mathrm{D}}=103$~meV.\cite{slack1964}
We note that the vibronic softening constant, as it is described in
Eq.~\ref{eq:kv}, only accounts for a single distortion mode. However,
for an accurate account of the pJT vibronic details of a defect in
a crystal, a continuum of fundamental and excited vibrational modes
would have to be included in the summation. Since this is not practical,
we leave a qualitative description supported on the agreement between
adiabatic calculations and the measurements. Hence, for temperatures
approaching 0~K, all zero-phonon vibrations obeying the selection
rules will contribute to the vibronic force constant $k_{\mathrm{v}}^{R}$.
If this contribution is not strong enough to produce a negative curvature
of the APES at $R=\mathrm{A}$, the vacancy will preserve the $|a_{1}^{1}\rangle\sim|3\bar{1}\bar{1}\bar{1}\rangle$
state. This could be the case for V$_{\mathrm{C}}^{+}(h)$ since there
is good agreement between the magnitude of the low-$T$ hyperfines
shouldering the EI6 EPR signal and those calculated for V$_{\mathrm{C}}^{+}(h,\mathrm{A})$.\cite{bockstedte2003,umeda2004b}
For instance, at $T=10$~K the Si$_{1}$ HF components parallel and
perpendicular to the crystallographic $c$-axis were measured as $A_{\Vert}=434$~MHz
and and $A_{\bot}=297$~MHz, respectively.\cite{umeda2004a} These
are to be compared with $A_{\Vert}=400$~MHz and and $A_{\bot}=275$~MHz
calculated for V$_{\mathrm{C}}^{+}(h,\mathrm{A})$.\cite{bockstedte2003}
With increasing temperature, additional phonon modes are populated,
and $k_{\mathrm{v}}^{R}$ becomes more negative. Should $k_{\mathrm{v}}^{R}<-k_{0}^{R}$,
the V$_{\mathrm{C}}^{+}(h,\mathrm{B})$ state $|a_{-}'^{1}\rangle$
will be lower in energy and the observed HFs should be converted to
those of the monoclinic defect. The calculations from Ref.~\onlinecite{bockstedte2003}
anticipate that this transformation would lead to a Si$_{1}$ HF splitting
with $A_{\Vert}=313$~MHz and $A_{\bot}=215$~MHz, which would explain
the strong and progressive decrease of the observed HF data to $A_{\Vert}=344$~MHz
and $A_{\bot}=237$~MHz at $T=293$~K.\cite{umeda2004a}

The observed trigonal pattern for the EI6 main signal and related
hyperfines between 5~K and room-temperature can also be explained
based on dynamic arguments. Around $T=5\textrm{-}10\,\mathrm{K}$
the HF signals arise from the $|a_{1}^{1}\rangle\sim|3\bar{1}\bar{1}\bar{1}\rangle$
state. This static trigonal state accounts for the observed $\sim40$\%
localization of the paramagnetic wave function on Si$_{1}$.\cite{umeda2004a}
On the other hand, at higher temperatures the distorted $|a_{-}'^{1}\rangle$
state is expected to quickly hop between all three equivalent states
$|11\bar{1}\bar{1}\rangle$, $|1\bar{1}1\bar{1}\rangle$ and $|\bar{1}11\bar{1}\rangle$.
Our calculated hopping barrier of 0.02~eV is compatible with this
behavior. Since the three states lead to approximately the same spin-density
amplitude and shape on all four Si radicals, the result is the observation
of two trigonal HF signals. One of them corresponds to Si$_{1}$,
while the other represents a shell of the three Si$_{2\textrm{-}4}$
atoms, also explaining the observed 1:3 amplitude ratio.\cite{umeda2004a}

The complex temperature-dependence of the V$_{\mathrm{C}}^{-}(k)$
EPR signal and related hyperfine peaks can also be discussed with
help of dynamic arguments. Below $T\approx40$~K the EPR data shows
a $C_{1h}$ symmetric defect with one pair of HF shoulders due to
mirror-symmetric Si$_{3,4}$ radicals. This is consistent with the
static D-structure with paramagnetic ground state $|a_{-}'^{2}\,a''^{1}\,a_{+}'\rangle\sim|00\bar{1}1\rangle$.\cite{trinh2013}
Increasing the temperature above 40~K leads to the quenching of the
Si$_{3,4}$ HF signal and the angular dependence of the main signal
acquires a trigonal shape. Based on the energy difference between
V$_{\mathrm{C}}^{-}(k,\mathrm{D})$ and V$_{\mathrm{C}}^{-}(k,\mathrm{C})$
(0.02~eV), as well as on the calculated rotation and transformation
barriers from Table~\ref{tab2} and Figure~\ref{fig8}, we suggest
that this effect is connected to the off-plane $\mathrm{D}\rightarrow\mathrm{C}'\rightarrow\mathrm{D}''$
transformation, which is limited by a barrier estimated as $E_{\mathrm{fwd}}=0.06$~eV.
Such a sequence effectively leads to the alternation between structures
C and D and to the rotation of the mirror plane of the defect. This
mechanism translates into alternate transitions between symmetry-equivalent
$|00\bar{1}1\rangle$ and $|\bar{1}100\rangle$ conjugate states from
structures D and C, respectively. The amplitude of the spin-density
on the basal radicals becomes \emph{intermittent}, explaining the
disappearance of these HFs from the V$_{\mathrm{C}}^{-}(k,\mathrm{D})$
spectrum at $T\approx40$~K. Above $T\approx40\textrm{-}80$~K,
new trigonal Si$_{1}$ and Si$_{2\textrm{-}4}$ HFs appear in the
EPR spectrum, which are assigned to the combined thermally activated
population and rotation of both V$_{\mathrm{C}}^{-}(k,\mathrm{D})$
and V$_{\mathrm{C}}^{-}(k,\mathrm{C})$ states.\cite{trinh2013} However,
this argument is incompatible with the above referred localization
intermittency which is expected to work above 80~K as well. We note
that unlike in $|a''^{1}\rangle$ of V$_{\mathrm{C}}^{-}(k,\mathrm{D})$,
the symmetry of the $|a_{+}'^{1}\rangle\sim|1\bar{1}00\rangle$ state
of V$_{\mathrm{C}}^{-}(k,\mathrm{C})$ does not impose zero amplitude
of the wave function at any of the Si nuclei. The spin-density on
Si$_{3}$ and Si$_{4}$ is actually small but not zero (see Figure~\ref{fig4}).
We suggest that for $T>40\textrm{-}80$~K the V$_{\mathrm{C}}^{-}(k,\mathrm{D})$
HFs are quenched due to intermittency effects, while V$_{\mathrm{C}}^{-}(k,\mathrm{C})$
becomes populated and its HF-related features increase in the spectrum.
During the thermally activated motion of V$_{\mathrm{C}}^{-}(k,\mathrm{C})$,
the Si nuclei in the basal plane always contribute with a non-zero
hyperfine interaction. This could explain the appearance of the weak
and broad Si$_{2\textrm{-}4}$ trigonal HF at about 80~K. V$_{\mathrm{C}}^{-}(k,\mathrm{C})$
with $|a_{-}'^{2}\,a_{+}'^{1}\,a''\rangle$ filling order is expected
to show a temperature-dependent pJT effect due increasing coupling
between $a'$ states. This would lead to the stabilization of this
state, making it consistent with the strong temperature dependence
of the Si$_{1}$ HF axial component, which increases from 28~MHz
at 60~K to 103~MHz at 140~K.\cite{trinh2013}

An identical argument can applied to V$_{\mathrm{C}}^{-}(h)$, which
adopts the ground state structure C. In this case, structure D is
metastable by 0.14~eV, the lowest $\mathrm{C}\rightarrow\mathrm{D}$
transformation barrier is 0.17~eV and V$_{\mathrm{C}}^{-}(h,\mathrm{D})$
does not appear in the spectrum. Below $T\approx60$~K the V$_{\mathrm{C}}^{-}(h,\mathrm{C})$
state is static ($|a_{+}'^{1}\rangle\sim|1\bar{1}00\rangle$) with
two inequivalent HFs on Si$_{1}$ and Si$_{2}$ on the symmetry plane.
Above $\sim70$~K the available thermal energy promotes the $\mathrm{C}\leftrightarrow\mathrm{C}'$
rotation mechanism (with calculated barrier of 0.08~eV), and the
spin-density intermittency on Si$_{2\textrm{-}4}$ nuclei results
in the observation of a single Si$_{1}$ axial HF. This was explained
in Ref.~\onlinecite{trinh2013}. Increasing further the temperature
above 120~K, led to the appearance of a trigonal Si$_{2\textrm{-}4}$
HF signal, which is in conflict with the intermittency argument. Again,
we suggest that raising the temperature also increases the magnitude
of the vibronic term for structure C, and that increases the amplitude
of the $|a_{+}'^{1}\rangle\sim|1\bar{1}00\rangle$ wave function on
Si$_{3}$ and Si$_{4}$ by further increasing the $|e'\rangle$ contribution
to $|a_{+}'\rangle$. The result is a dynamic state with amplitude
on all Si radicals, and a reconciliation of the model with the observations.

The reason for V$_{\mathrm{C}}^{-}(k)$ showing a nodal ($a''$) ground
state, as opposed to V$_{\mathrm{C}}^{-}(h)$ which is a symmetric
ground state ($a'$), is also due to the weaker crystal-field on site
$k$. As depicted in Figure~\ref{fig3}, for V$_{\mathrm{C}}^{-}(k)$
the Coulomb interaction between $|a_{-}'\rangle$ and $|a_{+}'\rangle$
states causes a crossing of levels between the neutral and the single
negatively charged states, and that results in a more stable $|a_{-}'^{2}\,a''^{1}\,a_{+}'\rangle$
filling order. Also in Figure~\ref{fig3}, it is evident that the
stabilization of the $|a_{-}'^{2}\,a_{+}'^{1}\,a''\rangle$ filling
on the $h$-site indicates that the repulsion between symmetric states
is weaker. Hence, the crossing effect observed in V$_{\mathrm{C}}^{-}(k)$,
is now obtained for the double negative charge state in V$_{\mathrm{C}}^{=}(h)$.

Finally, we would like to underline a fundamental issue regarding
the assignment of EH$_{6/7}$ to the superposition of negative-$U$
double donors from V$_{\mathrm{C}}(k)$ and V$_{\mathrm{C}}(h)$.\cite{booker2016}
If that was indeed the case, the peak amplitude of EH$_{6/7}$ would
have to match that of Z$_{1/2}$, and that is usually not observed
by DLTS. In fact, it is widely documented that EH$_{6/7}$ has a smaller
amplitude than Z$_{1/2}$, irrespective of the sample history, including
as-grown, irradiated/implanted plus annealing and thermal processed
(see for example Refs.~\onlinecite{danno2006,wongleung2008,alfieri2013,ayedh2014}).
This led the authors from Ref.~\onlinecite{ayedh2014} to the suggestion
that EH$_{6/7}$ is a single donor transition. Our calculations are
consistent with a broad EH$_{6/7}$ signal made of closely spaced
positive-$U$ ordered double donor levels of the vacancy in different
sub-lattice sites. Further support for the positive-U sequence of
donor levels come from the relative formation energies of V$_{\mathrm{C}}$
for different charge states. From these figures we can estimate the
fraction of positively charged vacancies, $f_{+}$ (with respect to
total amount of V$\mathrm{_{C}}$ defects) when the Fermi level is
located halfway between the $(0/+)$ and $(+/+\!+)$ levels,

\begin{equation}
f_{+}=\frac{\exp(U/2k_{\mathrm{B}}T)}{2+\exp(U/2k_{\mathrm{B}}T)},\label{eq:fraction}
\end{equation}
which for a negative-$U$ center means that the Fermi level is locked
at the $(0/+\!+)$ transition. Equation~\ref{eq:fraction} neglects
the formation of negatively charged vacancies, it assumes that V$_{\mathrm{C}}^{0}$,
V$_{\mathrm{C}}^{+}$, and V$_{\mathrm{C}}^{+\!+}$ states are in
thermal equilibrium at temperature $T$, and k$_{\mathrm{B}}$ is
the Boltzmann constant. Hence, at $T=5$~K and considering $U=-0.04$~eV
reported for EH$_{7}$ by Booker \emph{et~al.}\cite{booker2016}
we arrive at $f_{+}\sim10^{-21}$, essentially telling us that under
these conditions all vacancies would be EPR-inactive. This strongly
indicates that the real $U$-value of EH$_{7}$defect should be positive.
Analogous conclusions would be drawn for EH$_{6}$. 

\section{Conclusions}

We report on density functional calculations of the electronic and
dynamic properties of the carbon vacancy in $4H$-SiC using semi-local
and hybrid functionals. The defect exhibits a rich catalogue of structures
that depend on the sub-lattice site and charge state. Their occurrence
is rationalized on the basis of several effects, namely the character
of the occupied one-electron states, the site-dependence of the crystal-field,
and the magnitude of the pseudo-Jahn-Teller effect. Diamagnetic states
show either strong or no pseudo-Jahn-Teller effect and possess a relatively
deep potential energy surface. They display stable ground state structures
A, B and D for charge states $q=+2$, $0$ and $-2$, respectively,
no matter their sub-lattice site. Conversely, paramagnetic charge
states $q=+1$ and $q=-1$ suffer from weak pseudo-Jahn-Teller distortions.
Here, ground states are respectively B and D for the $k$-site, whereas
they are respectively A and C for the $h$-site. This structural variety
essentially arises from the stronger crystal-field on the $h$-site
that stabilizes electronic states polarized along the main axis. Also
for the paramagnetic states, and depending on the strength of the
vibronic coupling within the pseudo-Jahn-Teller effect, metastable
structures play an important role in the behavior of the vacancy as
a function of temperature. For charge states $q=+1$ and $q=-1$,
structures A and C are metastable in the sub-lattice site $k$, respectively,
while structures B and D are metastable in the sub-lattice site $h$.

Mechanisms for the transformation between these structures, as well
as for the rotation of the mirror plane for monoclinic structures
(B, C, and D) were calculated and discussed in the light of the temperature-dependence
of the EPR data. From the total energies, electrical levels and nudged
elastic band calculations we constructed a configuration coordinate
diagram which considers electronic transitions, as well as structure
rotations and transformations between the relevant configurations.

Regarding the electrical activity, our results support the assignment
of Z$_{1/2}$ and EH$_{6/7}$ DLTS signals to the acceptor and donor
transitions of the carbon vacancy. We were able to attribute a sub-lattice
site to each component of the DLTS signals based on (i) the correlation
between the relative magnitudes of the calculated and measured $U$-values,
and (ii) the correlation between the site-dependent formation energies
and the relative intensity of the DLTS peaks in as-grown material.
Accordingly, we support the assignment of Z$_{1}$ and Z$_{2}$ DLTS
peaks to $(=\!/0)$ two-electron cascade emissions from V$_{\mathrm{C}}^{=}(h)$
and V$_{\mathrm{C}}^{=}(k)$ defects, respectively. We were able to
apply analogous arguments to the donor transitions. In this case we
assign EH$_{6}$ and EH$_{7}$ peaks to electron emissions from V$_{\mathrm{C}}(h)$
and V$_{\mathrm{C}}(k)$ defects, respectively. Our results favor
a positive-$U$ ordering for the donor transitions in both sub-lattice
sites. Hence, each peak should result from $(0/+)$ and $(+/+\!+)$
transitions with very close emissions rates. Although this is at variance
with recent electrical measurements,\cite{booker2016} they are consistent
with the low-temperature EPR data acquired in darkness.

\section*{Acknowledgements}

We would like to thank Prof. Bengt Svensson for providing fruitful
comments and criticisms. This work was jointly supported by the Science
for Peace and Security NATO Program through project SPS 985215, and
by the \emph{Fundação para a Ciência e a Tecnologia} (FCT) through
project UID/CTM/50025/2013. The authors would like to further acknowledge
the computer resources provided by the Swedish National Infrastructure
for Computing (SNIC) at PDC.

\bibliographystyle{apsrev4-1}
\bibliography{refs}

\end{document}